\documentclass[conference]{IEEEtran}

\ifCLASSOPTIONcompsoc
  \usepackage[nocompress]{cite}
\else
  \usepackage{cite}
\fi

\usepackage{tikz}
\usepackage{amsmath}

\usepackage{amssymb}
\usepackage{booktabs}
\usepackage{multirow}
\usepackage{float}

\newtheorem{exmp}{Example}[section]

\newcommand{\user}{u}

\newcommand{\week}{D}
\newcommand{\dayy}{d} 

\newcommand{\raw}{\mathit{raw}}
\newcommand{\stat}{\mathit{stat}}
\newcommand{\dist}{\mathit{dist}}
\newcommand{\autoe}{\mathit{ae}}

\newcommand{\Stepdaily}{S}
\newcommand{\Stepdailyu}{\Stepdaily_u}

\newcommand{\step}{s}
\newcommand{\vecstep}{\vec{\step}}

\newcommand{\vecstepraw}{\vec{\step}_{\raw}}
\newcommand{\vecsteprawday}{\vec{\step}_{\raw}^{\,\dayy}}
\newcommand{\vecsteprawweek}{\vec{\step}_{\raw}^{\,\week}}
\newcommand{\vecsteprawuser}{\vec{\step}_{\raw, \user}}
\newcommand{\vecsteprawuserday}{\vec{\step}_{\raw, \user}^{\,\dayy}}

\newcommand{\vecstepstat}{\vec{\step}_{\stat}}

\newcommand{\vecstepdist}{\vec{\step}_{\dist}}
\newcommand{\vecstepdistuser}{\vecstep_{\dist,\user}}

\newcommand{\vecstepae}{\vec{\step}_{\autoe}}

\newcommand{\vecstepstatuserday}{\vec{\step}_{\mathit{stat}, \user} ^{\,\dayy}}

\newcommand{\vecstepuserday}{\vecstep_\user^{\,\dayy}}
\newcommand{\vecstepdailyu}{\vecstep_u}

\newcommand{\vecstepdailyv}{\vecstep_v}
\newcommand{\Link}{f^{\mathit{link}}}
\newcommand{\cosine}{\mathit{cos}}
\newcommand{\eucl}{\mathit{eucl}}
\newcommand{\densesiamese}{\mathtt{Dense\_siamese}}
\newcommand{\lstmsiamese}{\mathtt{LSTM\_siamese}}
\newcommand{\cnnsiamese}{\mathtt{CNN\_siamese}}
\newcommand{\bilstmsiamese}{\mathtt{biLSTM\_siamese}}
\newcommand{\attntnsiamese}{\mathtt{Attention\_siamese}}

\newcommand{\RFstandard}{\mathtt{RF\_standard}}
\newcommand{\unsupeucl}{\mathtt{Euclidean}}
\newcommand{\unsupcos}{\mathtt{Cosine}}

\newcommand{\windowsize}{w}
\newcommand{\bucketsize}{b}
\newcommand{\maxsteps}{\mathit{max\_steps}}

\newcommand{\layersize}{l}
\newcommand{\kernelsize}{k}
\newcommand{\maxpoolsize}{p}
\newcommand{\dropout}{\mathtt{d}}

\makeatletter
\newcommand{\linebreakand}{%
  \end{@IEEEauthorhalign}
  \hfill\mbox{}\par
  \mbox{}\hfill\begin{@IEEEauthorhalign}
}
\makeatother

\begin{document}

\title{You Are How You Walk: Quantifying Privacy Risks in Step Count Data}
\author{\IEEEauthorblockN{Bartlomiej Surma\textsuperscript{\textsection} and Tahleen Rahman\textsuperscript{\textsection}} 
\IEEEauthorblockA{CISPA Helmholtz Center for Information Security}
\{tahleen.rahman, bartlomiej.surma\}@cispa.saarland\\
\and
\IEEEauthorblockN{Monique Breteler}
\IEEEauthorblockA{German Center for Neurodegenerative Diseases (DZNE)}
monique.breteler@dzne.de\\
\linebreakand 
\IEEEauthorblockN{Michael Backes and Yang Zhang \footnote{Corresponding author}}
\IEEEauthorblockA{CISPA Helmholtz Center for Information Security}
\{backes, yang.zhang\}@cispa.saarland\\
}

\maketitle

\begingroup\renewcommand\thefootnote{\textsection}
\footnotetext{Equal contribution}
\endgroup
\begin{abstract}
Wearable devices have gained huge popularity in today's world. These devices collect large-scale health data from their users, such as heart rate and step count data, that is privacy sensitive, however it has not yet received the necessary attention in the academia. In this paper, we perform the first systematic study on quantifying privacy risks stemming from step count data. In particular, we propose two attacks including attribute inference for gender, age and education and temporal linkability. We demonstrate the severity of the privacy attacks by performing extensive evaluation on a real life dataset and derive key insights. We believe our results can serve as a step stone for deriving a privacy-preserving ecosystem for wearable devices in the future.
\end{abstract}

%
\IEEEpeerreviewmaketitle


\section{Introduction}
In the current era of the Internet of Things, extensive amounts of data are generated by users not just from internet browsing and social media but also smart devices like wearables, cars or even household appliances.
While some data is shared willingly and purposely by the users themselves (e.g. social media posts), some data is shared unknowingly (e.g. web browser telemetry, tracking pixels). 
Yet some other data, such as biomedical or genomic data is shared only with a trusted party for a specific
purpose (e.g. health situation monitoring and improvement, activity tracking and analysis, dietary and lifestyle advice/consultation).
This has led to huge interest in privacy concerns arising out of such data.
Recently, acts like the General Data Privacy Regulation or the California Consumer Privacy Act have imposed strict regulations for collecting, processing and sharing user data.

Certain types of data are classified as personal or even sensitive personal\footnote{https://www.burges-salmon.com/news-and-insight/legal-updates/gdpr-personal-data-and-sensitive-personal-data/}. 
However, many kinds of biomedical data (e.g. step counts, heart rate, SpO2, nutrition and sleep data) are in the gray zone, which are considered sensitive personal information only under special circumstances when they indicate diseases or disabilities.
In case of biomedical services, the user typically does not have fine grained control over the data shared, i.e., it is an all or nothing scenario.
Such data is extremely sensitive, as it can expose other information about the user, such as their medical condition, ethnicity, habits, kinship or financial status.
Therefore, biomedical data needs to be treated with extreme caution by the service provider.
Especially in the recent years, wearable devices have gained huge popularity and become inseparable parts of people's lives.
Data collected from these devices is used for a plethora of purposes by multiple parties, such as private companies, health organizations and government agencies.
On the other hand, this gives rise to increasing concerns over privacy of the user.

In two recent works \cite{jiang2019pedometer, jiang2020attr} Jiang et al. devise a new deep neural network model to predict age and gender from pedometer data. The authors use number of steps made each day collected over 259 consecutive days and are able to find different walking behaviors during week days, weekends and holidays. The shortcoming of these papers is that an attacker needs to collect data over hundreds of days, which is not very practical.

In this paper, we perform the first, extensive study of privacy risks arising from fine grained user’s step count data collected over short time spans such as one week or even a single day. Such data is collected by various activity trackers (e.g. Endomondo, FitBit, Apple’s Health), but can also be collected from the accelerometer on smartphones either by an app or an opened website. 
Such accelerometer data is not considered sensitive and therefore can be collected without explicit permission from the user. While concerns about privacy of walking patterns data were raised before \footnote{https://w3c.github.io/sensors/\#user-identifying}, to the best of our knowledge, a systematic in-depth analysis of the privacy issues has not been performed yet.

\subsection{Our Contributions}
We perform the first large-scale study of the privacy issues within fine grained user's step count data.
In particular, we design various attribute inference and linkability attacks. 
For the attribute inference attack, we assume that the adversary has access to the step count data of the target user for example collected from a smartphone.
The adversary then tries to infer certain personal attributes to which he normally would not have access.
For the linkability attack, the adversary has access to target user's step count data as well as an anonymized database containing further sensitive information along with the step count data collected at a certain point in time.
The adversary then tries to link the target user with the record in the database.
This could have a wide variety of implications like targeted advertisements, surveillance, unfair credit score and pricing by insurance companies or banks, to name a few.  
We demonstrate the performance of the attacks with an extensive evaluation on a real world dataset of 1000 participants. 

We make the following key contributions:
\begin{itemize}
\item In order to leverage the power of deep learning, we develop three conceptually different feature extraction methods, namely, statistical, distributions and autoencoders.

\item Our attribute inference attack is aimed at finding the correlation between the walking patterns of the user and three personal attributes namely gender, age and education. We find that gender and age can be inferred with a high confidence, while education does not show a strong correspondence with our data.

\item We also run an ensemble version of our attribute inference attack by splitting the user step data into actions based on active and non-active periods. We classify each action separately and infer the users' attribute based on all the actions of a user. 

\item We make a comparison between all the feature selection methods and classifiers on all the attribute inference tasks.

\item We run three different types on linkability attacks. The first one is unsupervised and it relies on the distance (under some metric) between feature vectors of different samples. The second attack uses a pairwise vector similarity metric between features of different samples to fit a traditional machine learning classifier, namely, random forest. The third one uses the One Shot Learning method with Siamese networks.

\end{itemize}

\subsection{Organization} The paper is organized as follows: in Section~\ref{relatedwork}, we present some related work. We introduce our dataset in Section~\ref{dataset}. In Section~\ref{sec:feature} we describe how we extract features from the raw data. Next in Section \ref{sec:attrinf_attack} we present our attribute inference attack followed by the experimental evaluation in Section~\ref{attr_inf_eva}. In Section~\ref{sec:link_attack} we present our user linkability attack, followed by its experimental evaluation in Section~\ref{sec:link_eva}. Finally we conclude our paper in Section~\ref{sec:conclusion}.
\section{Related Work} \label{relatedwork}

It has been shown, that it is possible to fingerprint Android devices by accessing the accelerometer API \cite{BojinovMNB14, Sanorita2014accelprint} from a website opened on the phone.
Bittel et al. \cite{bittel2015accuracy} shown that iPhone's accelerometer is just as accurate and precise as biomedical measurement devices.
Accelerometer data from a phone carried in a pants pocket is sufficient to predict owner's activity (e.g. sitting, standing, walking, running) \cite{guidoux2014activity, aguiar2014monitoring, pei2013behavior}. Accelerometer sensor data reveals how users hold and touch smartphones by hand, based on which Davarci et al. \cite{davarci2017age} perform child/adult detection.

Users' demographics inference has been studied in many previous works, which further confirms how important for privacy is such data. It has been demonstrated that users' gender and age can be inferred from their web browsing behaviors \cite{hu2007demographics}, mobile communication patterns \cite{dong2014demographics} or applications usage and web browsing on smartphones \cite{kalimeri2019demographics}.

The potential benefits of activity tracking for health and well-being has received a lot of attention from both academia and industry.
Shameli et al. \cite{shameli2017gamification} study how competitions affect physical activity using a dataset of competitions within the Argus smartphone app. 
Althoff et al. \cite{althoff2017large} study activity distribution from smartphones of 717,527 people across 111 countries.
Using the same dataset, Pierson et al. \cite{pierson2018modeling} use Cyclic Hidden Markov Models to detect and model activity cycles.

However, few have studied users’ behaviors when sharing such personal fitness data and the privacy concerns that arise from the collection, aggregation, and sharing of this data. 
Vitak et al. \cite{vitak2018privacy} highlight the relationship between users’ demographics, data sharing behaviors, privacy concerns, and internet skills.
Hassan et al. \cite{hassan2018analysis} demonstrate an attack against Endpoint Privacy Zones, that infers users’ protected locations from their information in public posts using activity data collected from the Strava app. 
Meteriz et al. \cite{meteriz2019you} predict the location trajectory of users from publicly available elevation profiles.
Nguyen et al. \cite{nguyen2019location} demonstrate high accuracy location tracking just through smartphone accelerometer and magnetometer's footprints.
\section{Dataset}\label{dataset}
Our step count dataset was collected by DZNE (German Center of Neurodegenerative Diseases) among inhabitants of a middle-size German city, as part of their Rheinland Study to identify factors that influence adult health over the lifespan and into old age. The participants gave their explicit consent to use their pseudonymized data for research purposes. Even pseudonymized, the data is very sensitive; we ensured it never leaves trusted devices and won't be able to share it with researchers from other institutes.

The data was collected using an ActivPal Sensor, a small device worn on the thigh , which was carried by users for 7 consecutive days.
This can, to a good extent, simulate step count data collected from any other wearable device (e.g. smartwatch) or from a phone kept in a pocket throughout the day.

\subsection{Data Description}
Our raw dataset consists of number of step in each 15s period, for 1000 participants. We exclude 3 users due to incomplete data. Each user has age, gender and education attributes. 

\begin{figure}[!t]
	\centering
	\includegraphics[width=\columnwidth]{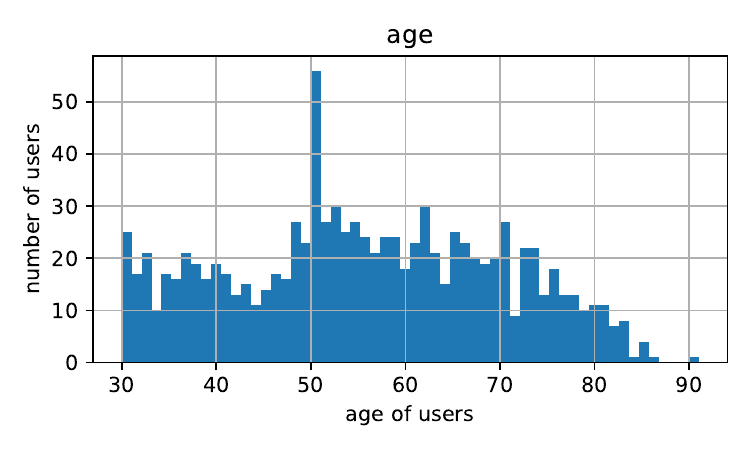}
	\caption{Distribution of age in our dataset \label{age}}
\end{figure}

\begin{table}[!t]
\centering
\begin{tabular}{c|l|l}
\toprule
\multicolumn{1}{l}{}       & Users   & 997     \\
\midrule
\multirow{2}{*}{gender}    & Males   & 44.13\% \\
                           & Females & 55.76\% \\
                           \midrule

\multirow{3}{*}{education} & High    & 52.35\% \\
                           & Medium  & 45.13\% \\
                           & Low     & 1.80\% \\
\bottomrule
\end{tabular}
\caption{Statistics of our dataset \label{tab:dataset}}
\end{table}

\begin{figure}[!t]
    \centering
    \includegraphics[width=\columnwidth]{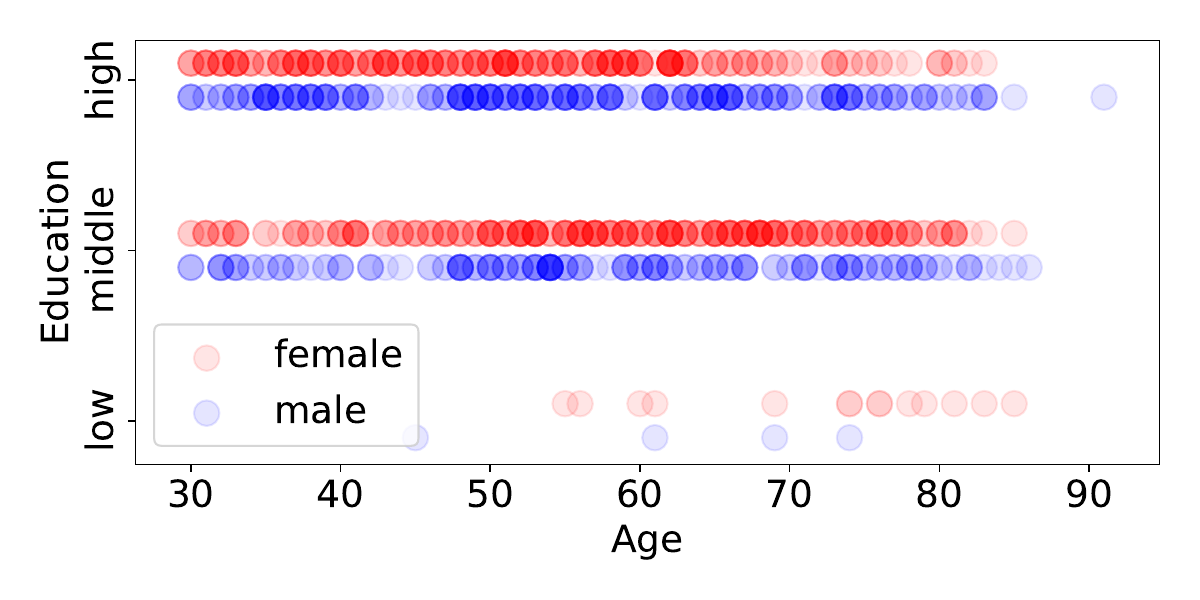}
    \caption{Distribution of users between all three attributes, the darker the color the more users with identical attributes \label{fig:attr_dist}}
\end{figure}

\begin{table}[!t]
    \centering
    \begin{tabular}{c|c|c}
    \toprule
         & age & education \\ \hline
        education & -0.209 & \\ \hline
        gender & -0.034 &  -0.130\\
\bottomrule        
    \end{tabular}
    \caption{Correlation coefficients between all three attributes of users \label{tab:attr_corr}}
    
\end{table}

\begin{figure}[!t]
	\centering
	\includegraphics[width=\columnwidth]{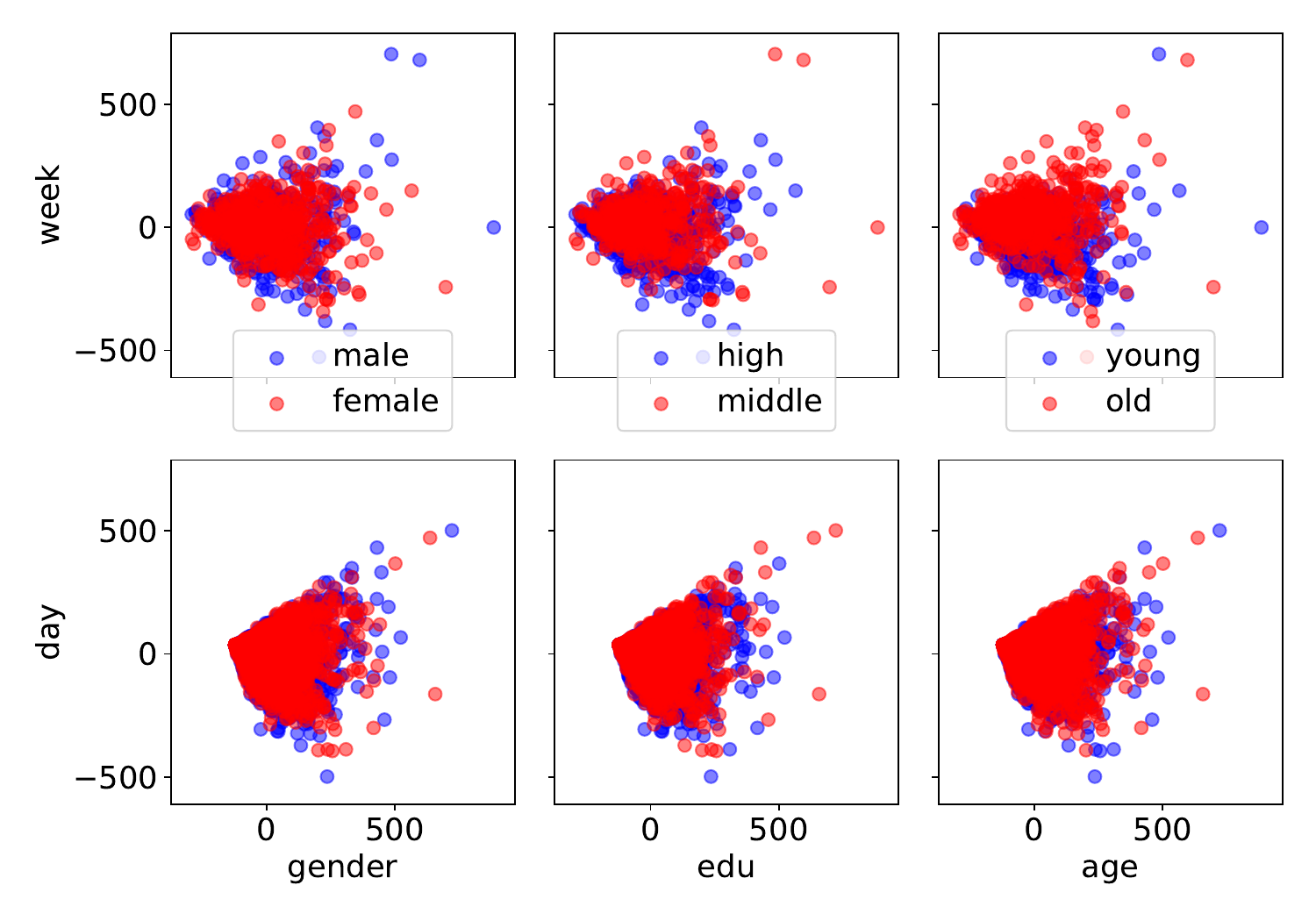}
	\caption{PCA of users raw step counts over the whole week (top) and split over 7 days (bottom) for each of our 3 attributes \label{pca}}
\end{figure}

The age of participants in our dataset ranges from 30 to 91 years old.
Figure \ref{age} shows the distribution of users of different ages in our dataset.
We divide users into two classes based on the median age (55 years). 

Table \ref{tab:dataset} shows the percentage of users in each class by gender and education.
For education there are 3 levels: 
\begin{itemize}
    \item \textbf{Low.} Early childhood education up to lower secondary education.
    \item \textbf{Middle.} Upper secondary education up to Bachelor degree or equivalent level.
    \item \textbf{High.} Master degree or equivalent level up to Doctoral degree or equivalent level.
\end{itemize}
For the classification task of education, we discard users with low education, since they comprise only $1.8\%$ of all participants.

Figure~\ref{fig:attr_dist} shows distribution of users among all 3 attributes. You can notice weak negative correlation of age and education. To measure it we encode education as $0, 1, 2$ for low, medium, high respectively and gender as 0 for male and 1 for female. Table~\ref{tab:attr_corr} confirms our observations.

Figure \ref{pca} shows the result of performing Principal Component Analysis (PCA) on users' raw step counts. 
The two colors denote the two classes for each attribute.  
We note that PCA on raw data does not suffice to separate users by gender, age or education.

Therefore, in the next section we present methods to preprocess this raw data in order to extract meaningful features for our tasks.

\subsection{Ethical Considerations}

We note that our primary institution does not provide an IRB nor mandate (or enable) approval for such experiments.
Meanwhile, the organization who collected the data followed the standard protocol for all the ethical compliance.
We also store the data in an independent server with f access control and during experiments, we strictly follow the data processing guideline defined by the data collecting organization.
\section{Feature Extraction} \label{sec:feature}

In this section we describe our three conceptually different feature extraction methods, followed by three different normalization techniques we use.

We refer to the sequence of the number of steps made by a user $\user$ every 15 seconds, as the raw step count vector, $\vecsteprawuser$.
These raw step count vectors $\vecsteprawuser$ are time series and thus are hard to work with when directly used as feature vectors for machine learning tasks.
Most standard machine learning techniques learn based on features that should be comparable between data points, which is not the case for time series.
A small displacement of an event in time would result in a feature vector much different from the original one (e.g., by cosine similarity).
For example if a user begins a morning walk even five minutes later than usual, the beginning and the end of the walk are now different features and this cannot be captured by traditional machine learning models.
To this end, we use different feature extraction methods as described next.

For a user $\user$, we collectively refer to the set of step count feature vectors from different days by $\Stepdailyu$, and a feature vector for a day $\dayy$ as $\vecstepuserday$. These $\vecstepuserday$ can either be the raw step counts or features created according to one of the methods below or their normalized versions.  
We omit subscript $\user$ when it is clear that we are talking about one specific user. 
Similarly, we omit the superscript $\dayy$ when the $\dayy$ is irrelevant.

\subsection{Statistical Method}\label{stats_features}
For each user, we first split $\vecstepraw$ into smaller, non overlapping subvectors of a defined window size $\windowsize$ (the last subvector might be smaller).
Then we calculate basic statistical values, which reflect different characteristics of a user's walking behavior: sum (how much the user walks), maximum (highest speed of the user), mean (how fast the user walked on average), median (around what speed did user use most of the time) and standard deviation (how much acceleration / deceleration did user do) on each subvector and combine all the calculated values in a single output vector $\vecstepstat$.
We use different subsets of statistics and different values of $\windowsize$ in our experiments. 

\begin{exmp}
Let $\vecstepraw = (5, 0, 0, 2, 3, 4, 3, 0), \windowsize = 3$, $\stat =\{ \mathit{sum},\mathit{mean} \} $. We split $\vecstepraw$ into three subvectors $(5, 0, 0)$,  $( 2, 3, 4)$ and $ ( 3, 0)$. 
Thus $ \vecstep_{\{ \mathit{sum},\mathit{mean} \} } =$   
$ (\mathit{sum}(5, 0, 0) , \mathit{mean}(5,0,0), 
\mathit{sum}(2,3,4) , \mathit{mean}(2,3,4),$ 
$
\mathit{sum}(3,0) , \mathit{mean}(3,0) )
=(5, 1.67, 9,  3, 3, 1.5) $.
\end{exmp}

\subsection{Distributional Method}\label{dist_features}
We want to capture a distribution of steps done in every 15s period.
We also want to retain the information about time of a day when users walk more or less. 
Therefore, we split the $\vecstepraw$ into subvectors, calculate distributions on them and finally concatenate them.

Precisely, we find the maximal number of steps $\maxsteps$ any user makes in 15s. Then, for each user, we split $\vecstepraw$ into smaller, non overlapping subvectors of a defined window size $\windowsize$ (the last subvector might be smaller). 
We group all possible number of steps $0, 1, 2, \dots, \maxsteps$ into buckets of size $\bucketsize$.
Then for each resulting vector, we count the number of occurrences of steps in each bucket.
Since $0$ steps is the most common event, we have a bucket containing just $0$ steps, and further buckets containing $\bucketsize$ different number of steps.
We combine the calculated value counts into a single output vector $\vecstepdist$. 

\begin{exmp}
For $\vecstepraw = (5, 0, 0, 2, 3, 4, 3, 0)$, $\windowsize = 3$, $\bucketsize = 3$, $\maxsteps = 6$. We split $\vecstepraw$ into three subvectors $(5, 0, 0)$,  $( 2, 3, 4)$ and $ ( 3, 0)$. Each subvector will have three buckets, i.e., $\{0\}, [1,3] ,$ and $[4,6]$. 
For the first subvector $(5, 0, 0)$, we have $2$ instances in the first bucket, i.e., $\{0\}$, $0$ instances in the second bucket, i.e., $[1,3]$ and $1$ instance in the third, i.e, $[4,6]$. Analogous calculations are done for the next two subvectors.
The resulting feature vector is then $\vecstepdist = (2, 0, 1, 0, 2, 1, 1, 1, 0)$.
\end{exmp}

\subsection{Autoencoder}
\begin{figure}[!t]
	\centering
	\includegraphics[width=\columnwidth]{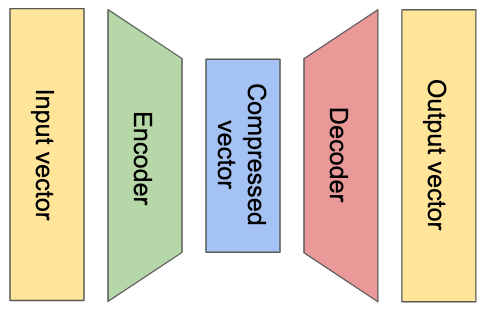}
	\caption{Illustration of a basic Autoencoder\label{fig_autoencoder}}
\end{figure}

Autoencoders are a type of neural networks consisting of an encoder and decoder.
In general, the encoder compresses the input into a lower dimensional vector and the decoder recreates the original input from it.
The unsupervised training objective of the network is to reconstruct the original input with as little loss as possible.
Because of the created ``bottleneck'', a data compression method is learnt by the neural network.

Once trained, the autoencoder model can be used to transform data into their compressed representations by only using the encoder part.
Autoencoders are widely used for denoising, anomaly detection, feature extraction and more.

In this work,
we first train an autoencoder with the normalized $\vecstepraw$ of all the users in our training set to train the model.
We then encode each vector into a compressed representation $\vecstepae$ with the trained encoder part of our network. We then use the resulting $\vecstepae$ as feature vectors for attacks.
We create two variants of autoencoders: one consisting of densely connected layers and the other one consisting of 1 dimensional Convolutional Neural Network (1D CNN) layers.

\subsection{Normalization} \label{data_norm}
We now describe the normalization techniques we use on the raw data as well as the features generated by the methods above.
\subsubsection{Feature-wise Normalization}\label{norm} In this method, each $i$-th value of a feature vector is divided by the maximum of $i$-th values of all feature vectors. This normalization is mostly useful for statistics and distribution features.

\subsubsection{Vector-wise Normalization} In this method, each element of an input vector is divided by the value of the largest element of this vector. For raw step count data, such processing removes information of how fast a user can walk, but keeps information when does she walk more and when less.

\subsubsection{Probability Distribution} In this method, each element of a vector is divided by the sum of all of its elements. Such processing preserves information about the overall relative walking speed of a person, but loses information about the total amount of steps made during the concerned time period.

\section{Attribute Inference Attack}\label{sec:attrinf_attack}
It has been shown that social network user’s personal attributes, e.g., gender, age, political believes, location, occupation, even if not disclosed, can be inferred from their friends \cite{al2012homophily}, online behaviours \cite{gl2018attribute} or from the content of their messages \cite{ciot2013gender}. In this work we study whether gender, age and education can be inferred from user’s step count data itself.

Using a wide range of feature vectors extracted from raw step data, we train multiple different machine learning and deep neural network models on the attribute classification task and report the results in Section~\ref{attr_inf_eva}.

\subsection{Experimental Setup}
Out of 1000 users in our dataset we need to filter out 3 of them due to step measurement errors.
We focus on binary classification of 3 attributes: gender, age and education.
For gender, we classify between male and female, where 56\% of participants are female.
For education, we focus on middle and high level education as explained before.
Age of the participants range from 30 to 91 years old, we choose the median (55 years) and classify participants as younger than 55, or 55 and older, giving us 50\% of users in each group.

We assume an adversary that has access to some amount of user's data with their attributes for training her models and unlabeled step count data of users whose attributes she wants to infer.
To simulate this settings, for each attribute, we randomly choose 80\% of users of each class as training data set and the remaining 20\% as testing data set.

We run an extensive amounts of attacks for attribute inference. We first use a wide range of feature extraction techniques on the raw dataset, and then train and test many different classifiers on those feature vectors.

\subsection{Feature Extraction}
We will show our methods for feature extraction in a step by step manner.
\begin{enumerate}
    \item\label{feature:split} The raw step vector of each user $\vecstepraw$ we take as a whole week $\vecsteprawweek$ or split it into different days, resulting in $7$ $\vecsteprawday$ for each user. This results in $2$ feature vector types.
    \item\label{feature:stat} We run statistical methods (as described in \ref{stats_features}) for all possible subsets of $\{\mathit{max}, \mathit{mean}, \mathit{median}, \mathit{std}, \mathit{sum}\}$ (other then $\varnothing$) for window sizes $\windowsize \in \{12, 24, 48, 60, 120, 240, 480, 720, 960, 1440, 1920,$ $2880, 5760\}$ on $\vecsteprawday$ and for window sizes $\windowsize \in \{240, 480, 720, 960, 1440, 1920, 2880, 5760, 40320\}$ on $\vecsteprawweek$. 
    This results in $(2^5 - 1) * (13 + 9) = 682$ feature vector types.
    \item\label{feature:dist} We run distributions as described in \ref{dist_features} on $\vecsteprawweek$ and $\vecsteprawday$ for window sizes $\windowsize \in \{240, 720, 1440, 2880\}$ and for bucket sizes $\bucketsize \in \{2, 4, 8\}$.
    This results in $2 * 4 * 3 = 24$ feature vector types.
    \item We apply three different normalization methods as described in \ref{data_norm} to all feature vector types we made so far. This results in $(2 + 682 + 24) * 3 = 2124$ normalized feature vector types.
    \item On each normalized feature vector type we train a densely connected auto encoder. Because some of them are much shorter then others we need to adjust the layer sizes to avoid compressing a shorter vector into a single value.
    The densely connected auto encoder consists of five densely connected layers of sizes $(\layersize_1, \layersize_2, \layersize_3, \layersize_4, \layersize_5)$,
    where $\layersize_1$ is equal to the length of the input feature vector,
    \begin{align*}
    \layersize_2 &= 
    \begin{cases}
        \mathit{min}(2048, \lfloor\layersize_1 / 4\rfloor), & \text{if } \layersize_1 > 255\\
        \lfloor\layersize_1 / 2\rfloor,              & \text{otherwise,}
    \end{cases}\\
    \layersize_3 &= 
    \begin{cases}
        \lfloor\layersize_2 / 4\rfloor, & \text{if } \layersize_2 > 127\\
        \lfloor\layersize_2 / 2\rfloor,              & \text{otherwise,}
    \end{cases}
    \end{align*}
    $\layersize_4 = \layersize_2$, $\layersize_5 = \layersize_1$.
    We then take third layer and use it as a new feature vector, which results in 2124 feature vector types.
    \item On each normalized feature vector type we train a convolution auto encoder.
    It consists of two convolution layers with 8 filters each, $\kernelsize_1$ and $\kernelsize_2$ kernel sizes respectively and a max pooling layer in between with the pool size of $\maxpoolsize$. Then two transposed convolution layers with an unpooling layer between them with corresponding parameters follow. Because of how long it takes to train a cnn autoencoder, we use the following sets of parameters $(\kernelsize_1, \kernelsize_2) \in \{(6, 6), (9, 6), (9, 9), (21, 9)\}$ and $\maxpoolsize \in \{2, 4\}$ only on weekly feature vectors $\vecsteprawweek$ normalized with feature-wise normalization and $\kernelsize_1 = 21, \kernelsize_2 = 9, \maxpoolsize = 4$ for the rest. This results in $4 * 2 * 1 + 2121 = 2129$ feature vector types.
    \item Additionally, we use a third type of data splitting from step \ref{feature:split}, that we call actions.
    We take a whole week of step data and extract all periods of maximal length that have no subperiod of 8 consecutive periods with 0 steps done (2min of rest) and use each such action together with the length, and time of beginning of the action as a feature vector.
    We then apply statistics (all statistics together) from step \ref{feature:stat} and distributions from step \ref{feature:dist} with the window size equal to action length to obtain 3 (together with raw) feature vector types. The reasoning behind it is that maybe a combination of shorter activities are better suited to predict person's attributes.
\end{enumerate}

In total we are testing 7085 feature vector types of either seven or one vector per user and 3 actions feature vector types.

\subsection{Classifiers}
We classify on three different binary tasks: gender (male or female), education (medium or high) and age (below 55 or above). We use the following classifiers from sklearn on each task with each feature vector type: Random Forest, Linear Regression, Support Vector Machine, Linear Support Vector Machine and 3 Densely Connected Neural Networks, all consisting of a densely connected and dropout layers. If $\layersize$ is the feature vector length and $\dropout$ is a 20\% dropout layer, then the networks will look as follows: $(\layersize, \frac{1}{4}\layersize, \dropout, 1), (\layersize, \frac{1}{2}\layersize, \dropout, \frac{1}{8} \layersize, 1), (\layersize, \frac{1}{2}\layersize, \dropout, \frac{1}{4}\layersize, \dropout, \frac{1}{16}\layersize, 1)$.

Now each feature vector set is split into 80\% training and 20\% testing data in a balanced way (e.g., if the classification is to be performed for gender, the 80 - 20 split is done on male and female separately and only then combined and shuffled together for both training and testing). For the feature vector sets, where we have 7 vectors per user, we make sure, that all the vectors are in either training or testing set for each user.

In addition for actions feature vectors an additional aggregation step is needed, because our objective is to classify an attribute of a person and not the action itself.
After obtaining scores from the classifier for each action of a specific person we discard 50\% of the results, that are the least sure about the attribute and then calculate both arithmetic mean and majority vote.

Due to long training time, we run CNN classifiers only on selected feature sets, namely raw daily and weekly step count with both feature-wise vector-wise normalization, all 3 actions, 4 distributions and 4 statistics (2 best performing on average and 2 with highest performance, as measured on other classifiers).
Each CNN classifier looks as follows: 2 convolution layers with 16 filters each, $\kernelsize_1$ and $\kernelsize_2$ kernel sizes respectively, 50\% dropout layer, max pooling layer with the pool size of $\maxpoolsize$, fully connected layer with 100 neurons and a fully connected layer with a single perceptron.
We use $(\kernelsize_1, \kernelsize_2) \in ((21, 9), (6, 6))$ and $\maxpoolsize \in (2, 4, 8)$

We run 3 LSTM classifier types, regular LSTM, bidirectional LSTM and bidirectional LSTM with attention. We use a Luong-style dot-product attention layer, \cite{luong2015effective}.  We only run LSTMs on feature vectors resulting from fixed window size aggregations, i.e. statistical (with window size $\windowsize \in \{60, 240, 720\}$), distributional (with window size $\windowsize \in \{60, 240, 720\}$, bucket size $\bucketsize \in \{2, 4\}$, statistical activities (with all the statistics combined) and distributional activities (with bucket size $\bucketsize = 2$). We split the feature vectors according to actions or windows and feed each subvector one by one to the LSTM. With an analogy to training an LSTM on a sentence of words, where each word is encoded in a fixed size vector, each window or activity is a separate word and the result of distribution or statistics function is the embedding. All three LSTMs consist of one respecive LSTM layer with 16 or 32 units, one dropout layer with 0.2 dropout rate and a fully connected layer with a single perceptron.
\section{Attribute Inference Evaluation} \label{attr_inf_eva}
Our goal was not to train the best possible classifier for the attribute inference task, but rather to cover a wide range of different classifiers and feature vectors.
We use AUC (Area Under ROC Curve) for measuring the performance of our classifiers.
As a result, we have 149433 classification results in total: 7 non-CNN classifiers for 3 different attribute types run on 7085 feature vector types, 3 actions feature vector sets with either mean or majority vote, 6 CNN classifiers for 3 different attribute types run on 12 selected feature vector types and 3 actions feature vector sets with either mean or majority vote and 6 LSTM classifiers 3 different attribute types run on 9 selected feature vector types and 2 action feature vector sets.

\subsection{Normalization Method}
The normalization methods do not have a strong influence on the results. On average the best performing is feature-wise normalization, with the exception of random forest classifier, where normalization is not needed and it would, in almost all cases, cause a small drop in performance.

\subsection{Statistical Method}
\begin{figure}[!ht]
	\centering
	\includegraphics[width=\columnwidth]{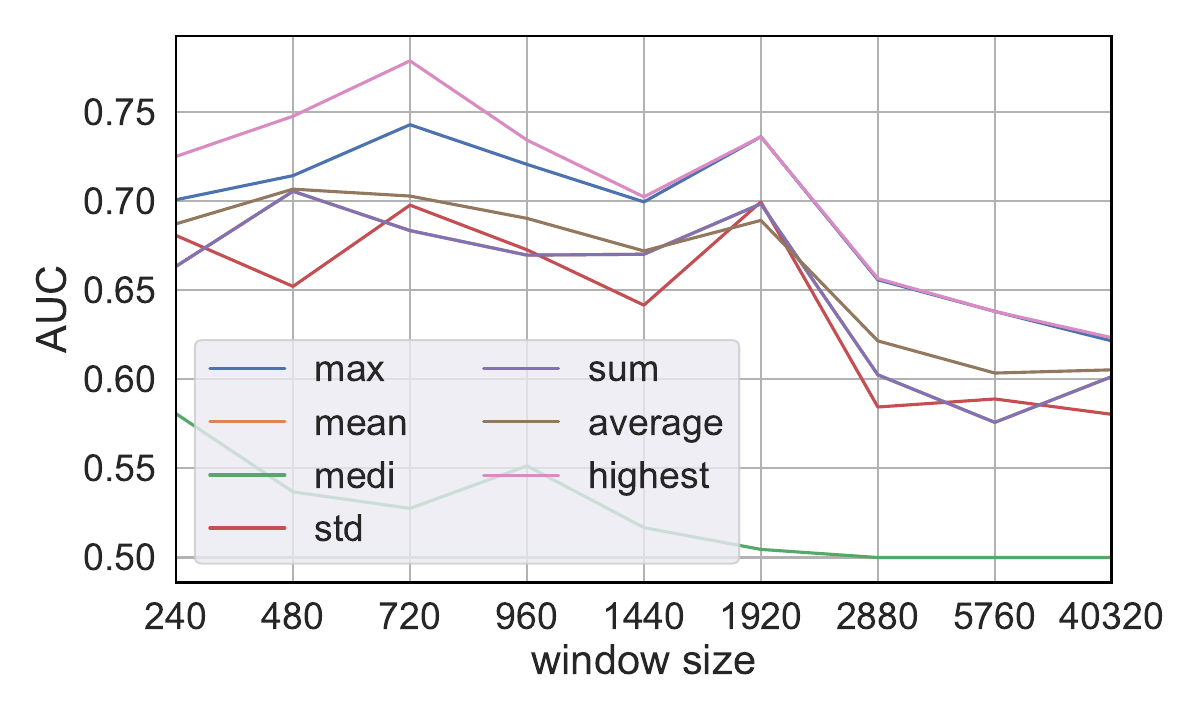}
	\caption{Influence of window size on age prediction \label{fig:stat_window}}
\end{figure}

We first look at the best window size for the statistical methods. On Fig.~\ref{fig:stat_window} we can see best results for single statistics on different windows on the task of predicting age. Additionally we also plot highest and average result of all the (combined) statistics. We can see that the best results are obtained for $\windowsize = 720$ and $\windowsize = 1920$, which are 3 and 8 hours periods correspondingly. The best performing single statistic is maximal number of steps and the worst is median. Sum and mean of the steps are overlapping on the plot, since mean is just sum divided by a constant. Especially interesting is the 8 hours period since it reflects differences in people's habits during the night (from midnight to 8 a.m.), working period (from 8 a.m. to 4 p.m.) and leisure time (4 p.m. to midnight). With only the maximal number of steps done in each period we can already infer person's age with the AUC score of 0.74 with a linear regression classifier.

\begin{figure*}[!ht]
	\centering
	\includegraphics[width=0.9\textwidth]{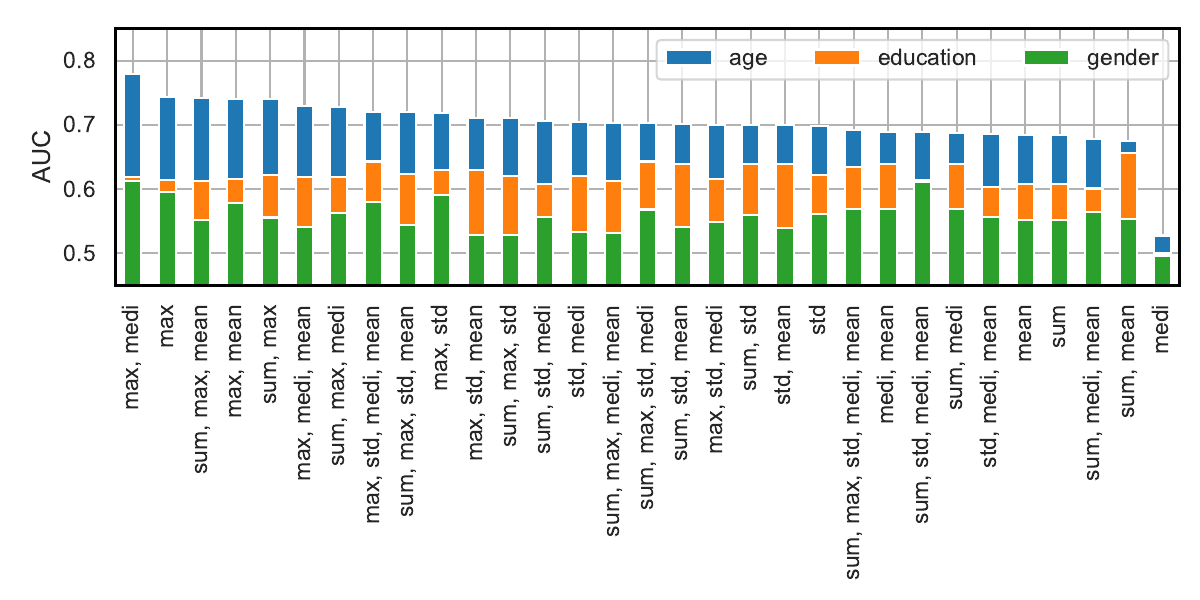}
	\caption{Influence of statistics subset on age predictions \label{fig:stat_subset}}
\end{figure*}

On Fig.~\ref{fig:stat_subset} we analyze best performing subsets of statistics on the window size $\windowsize = 720$. We can see that the easiest classification task is to predict user's age, then education and the hardest one is gender. Interestingly the best performing for both gender and age prediction is a combination of maximal and median values - the best and worst performing when taken alone. In general, combinations of more then two statistic values do not perform well. When a classifier has too many features and not enough data to train on, its performance will usually be lower then with less but well chosen features.

For all attributes the best results are found on week long step count series.
The best performing for age was a random forest classifier trained on maximal and median values on windows of size $\windowsize = 720$ (3 hours), achieving 0.78 AUC.
The best result for education is achieved by the smallest densely connected neural network classifier (with only 1 hidden layer) trained on maximal value of windows of size $\windowsize = 240$ (1 hour) with feature-wise normalization, achieving 0.68 AUC.
The best performance for gender was a random forest classifier trained on maximal values of windows of size $\windowsize = 1440$ (6 hours), achieving 0.65 AUC.
We can see that for age and gender the best discriminator is how fast and how fast on average are people walking, while for education it is more important how much are people walking. We are confident with prediction regarding age, but much less so when it comes to education or gender.

\subsection{Distributional Method}
\begin{figure*}[!ht]
	\centering
	\includegraphics[width=\textwidth]{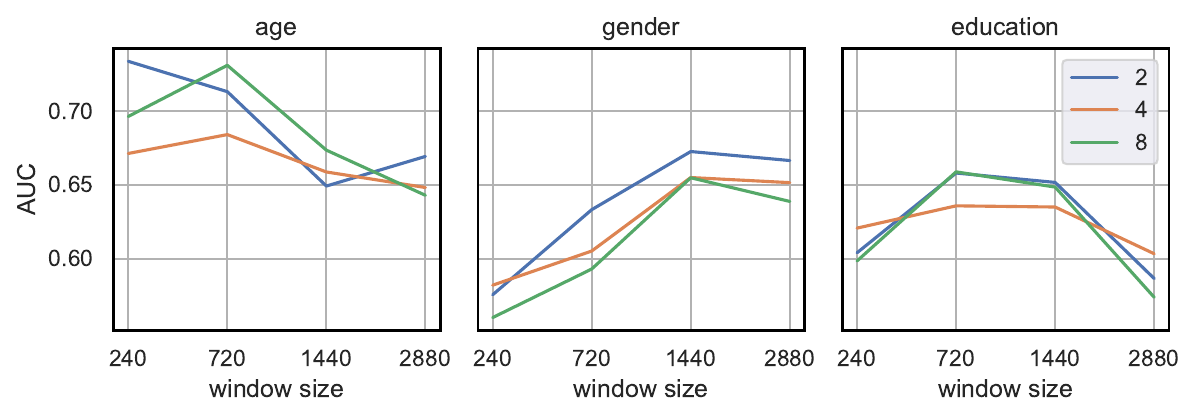}
	\caption{Influence of the window and bucket size for distributional method on the prediction quality \label{fig:dist_param}}
\end{figure*}

On Fig.~\ref{fig:dist_param} we can see best results for all bucket sizes, window sizes and target attributes.
We can observe, that bucket size $\bucketsize = 2$ has always the best (or almost best) performance, meaning more fine grained distribution has a positive impact on the results (even though it generates more features that classifier will need to learn how to process).
The best result for each attribute are as follows:
\begin{itemize}
    \item\textbf{Age.} Random forest classifier trained on $\bucketsize = 2$ and $\windowsize = 240$ (1 hour) achieves 0.73 AUC,
    \item\textbf{Gender.} Support vector classifier trained on $\bucketsize = 2$ and $\windowsize = 1440$ (6 hours) achieves 0.67 AUC,
    \item\textbf{Education.} Densely connected neural network (with 2 hidden layers) trained on $\bucketsize = 8$ and $\windowsize = 720$ (3 hours) achieves 0.66 AUC.
\end{itemize}

\subsection{Autoencoders}
We observe that autoencoders almost always cause a drop in AUC as compared to the feature vectors on which they were trained.
The best autoencoder was a densly connected auto encoder on weekly max and mean statistics on windows of size 960 with feature-wise normalization, age classificiation logistic regression achieved 0.77 AUC value, while the same classifier with the same feature vectors but without auto encoder step achieved 0.73 AUC.
The improvement, although non-negligable, is an exception rather than a trend.
It might be the case, that we do not have enough data for the autoencoders to efficiently learn an intermediate representations, and training directly for the task of prediction show better results.
It might, however, still be a useful technique if we have access to big number of unlabeled step count timeseries and we only know attributes of a small percentage of them.

\subsection{Actions}
On average arithmetical mean outperforms majority voting of the result scores, by a small margin. However in some of the best performing cases, mean can have a drastic improvement (e.g., for plain steps and a CNN gender classifier, after cross validation, majority vote achieves 0.53 AUC, while mean gives 0.78).
The best results were obtained for gender prediction with raw actions and a random forest classifier resulting in AUC of 0.87.
For age the best was a densely connected neural network classifier (with 2 hidden layers) trained on distribution of actions and achieved AUC of 0.69.
For education the best was logistic regression trained on distributions of actions achieving AUC of 0.61.
While with all other methods, gender was difficult to predict, actions can do it with a very high confidence. It suggests that there exists certain walking patterns characteristic for either man or woman.

\subsection{LSTMs}
Because LSTMs use feature vectors as a 2d matrix rather then a long vector we do not include them in the previous evaluations of the feature selection methods, but instead dedicate them this section.
For gender and education, distributional method outperforms statistical ones, while for age the statistical gives best results.
Splitting the data based on activities rather then a fixed window size causes a big drop in the AUC score for all the attributes. For both gender and age, window size $\windowsize = 240$ gives the best result. For distributional methods, the bucket size does not seem to have a big influence on the AUC score, but best results are achieved for $\bucketsize = 2$.
Making the LSTM bidirectional or adding attention both improve the results for gender prediction, but do not achieve higher performance on the other two attributes. It is possible that due to larger amount of trainable weights, more data would be needed to make use these additions to the basic LSTM.

\subsection{Summary}
Because of the amount of classifiers trained, we were not able to cross validate each experiment and rather we take the best performing ones (described in details in previous subsections) and run them again with 5 fold cross validation.
This approach results in us reporting potentially lower best performance values then possible, since we might have missed some well performing methods due to unlucky split into training and testing data, while ensuring, that we do not report high, lucky performance scores.

In general, weekly walking patterns are much better then daily for the attribute prediction task. Daily patterns can differ drastically between weekdays and weekends and thus, looking at a person's whole week is advised.

\begin{figure}[!ht]
    \centering
    \includegraphics[width=0.9\columnwidth]{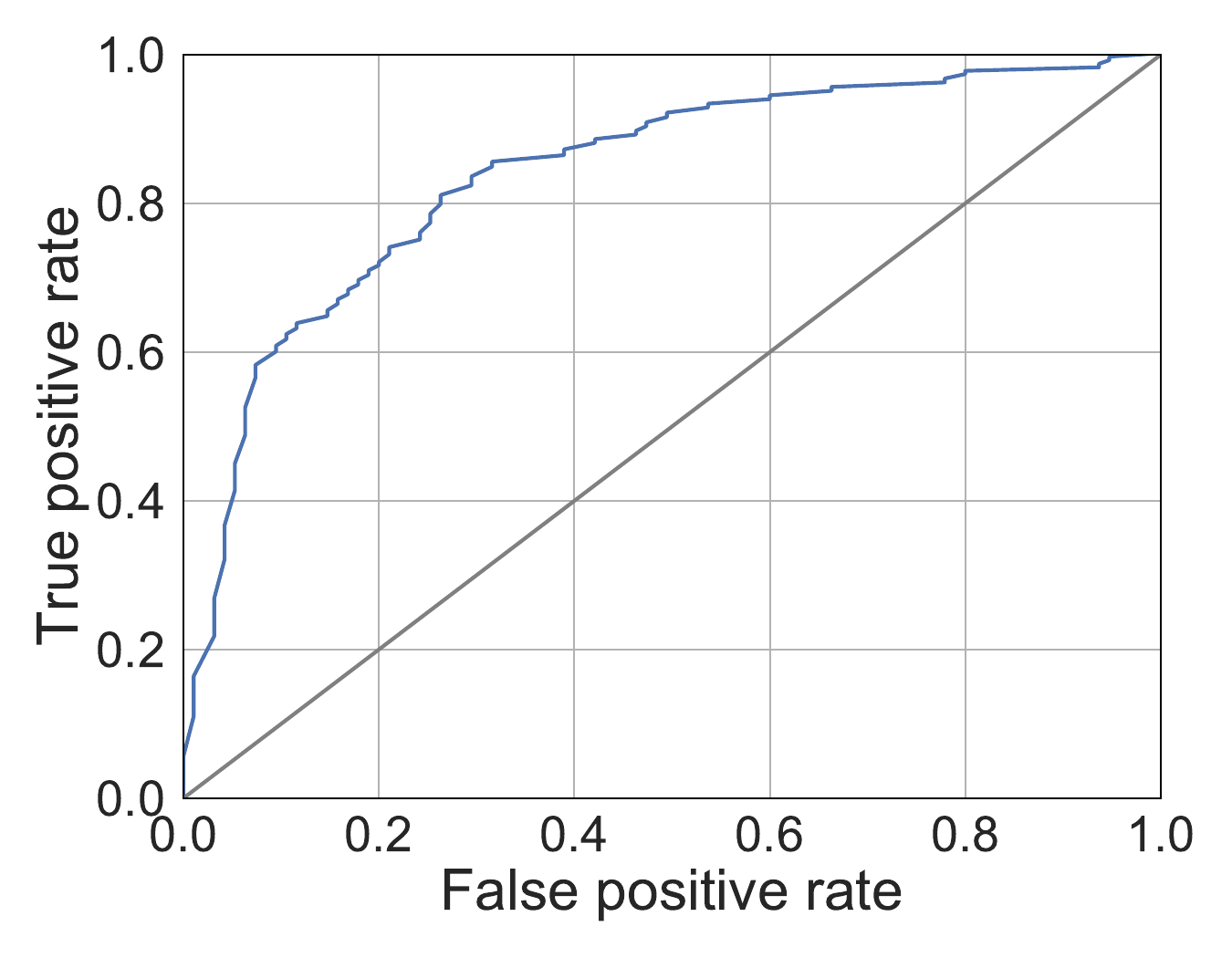}
    \caption{ROC curve for best performing age classifier (in blue). The gray line indicates the baseline (random guess). \label{fig:age_roc}}
\end{figure}

\begin{figure}[!b]
	\centering
	\includegraphics[width=\columnwidth]{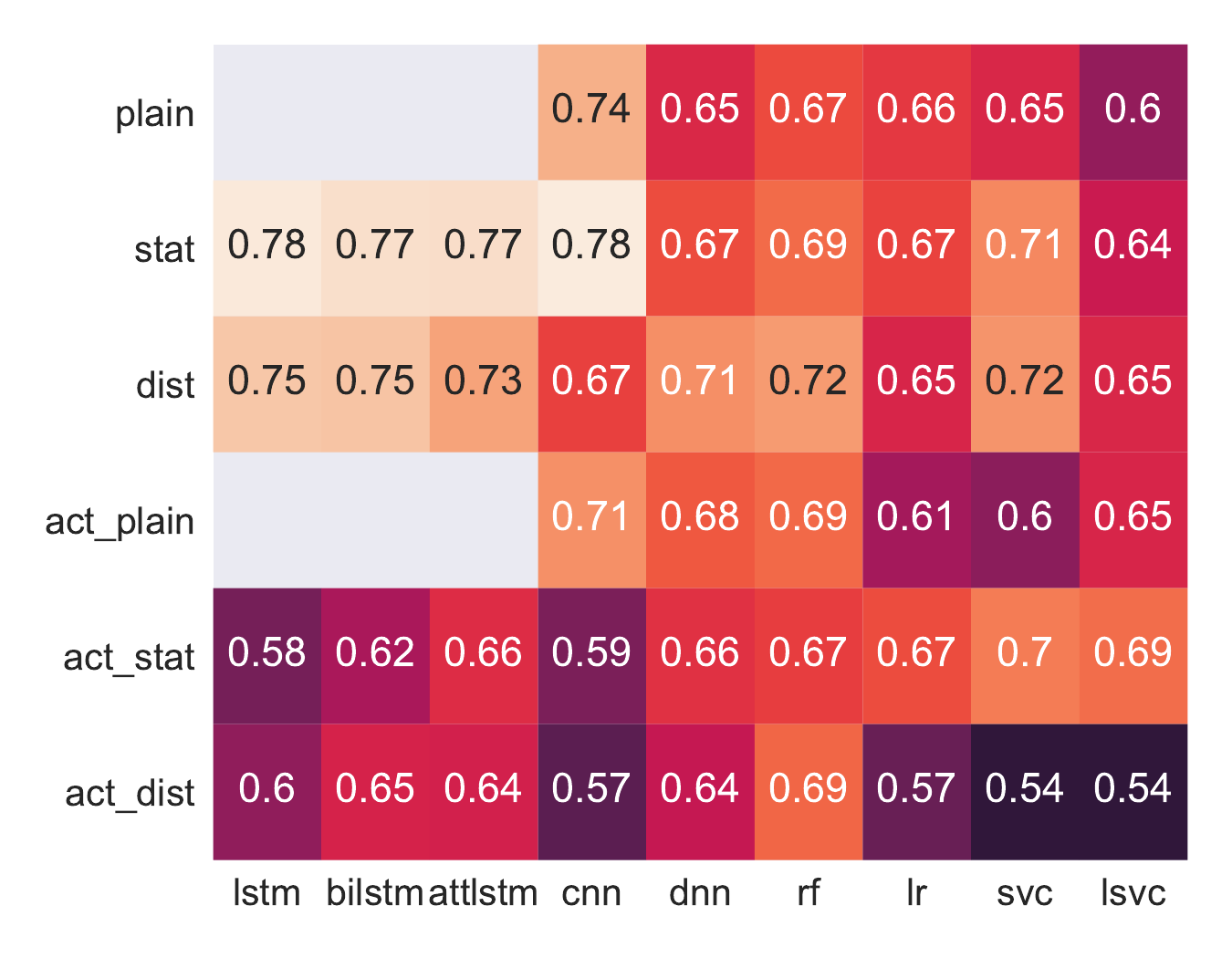}
	\caption{Best AUC scores for age prediction \label{fig:age_heat}}
\end{figure}

The result for age prediction are in Fig.~\ref{fig:age_heat}. It is the easiest attribute to predict and many classifiers achieve AUC greater then 0.7. Great performance of statistical feature vector generation confirms our observations, that for age inference the most important is how fast (in steps per minute) a person walks, as older people tend to walk slower. Data split into activities instead of fixed size windows does not improve the predictions, which means that no activities were found that would clearly indicate person's age.
LSTMs perform very well on both statistical and distributional methods, however the best result is achieved by a CNN consisting of 2 convolution layers with 6 filters of size 16 each, a 0.5 dropout layer, max pooling layer of pool size 8, dense layer with 100 perceptrons and finally dense layer with 1 perceptron, run on a single maximum statistic on window of size 240 (maximum number of steps done every hour). The average AUC over 5 fold validated training-test split is 0.778 with standard deviation of 0.047. In Fig.~\ref{fig:age_roc} we show the Receiver Operating Characteristic curve for the best cross validation run. 

\begin{figure}[!b]
	\centering
	\includegraphics[width=\columnwidth]{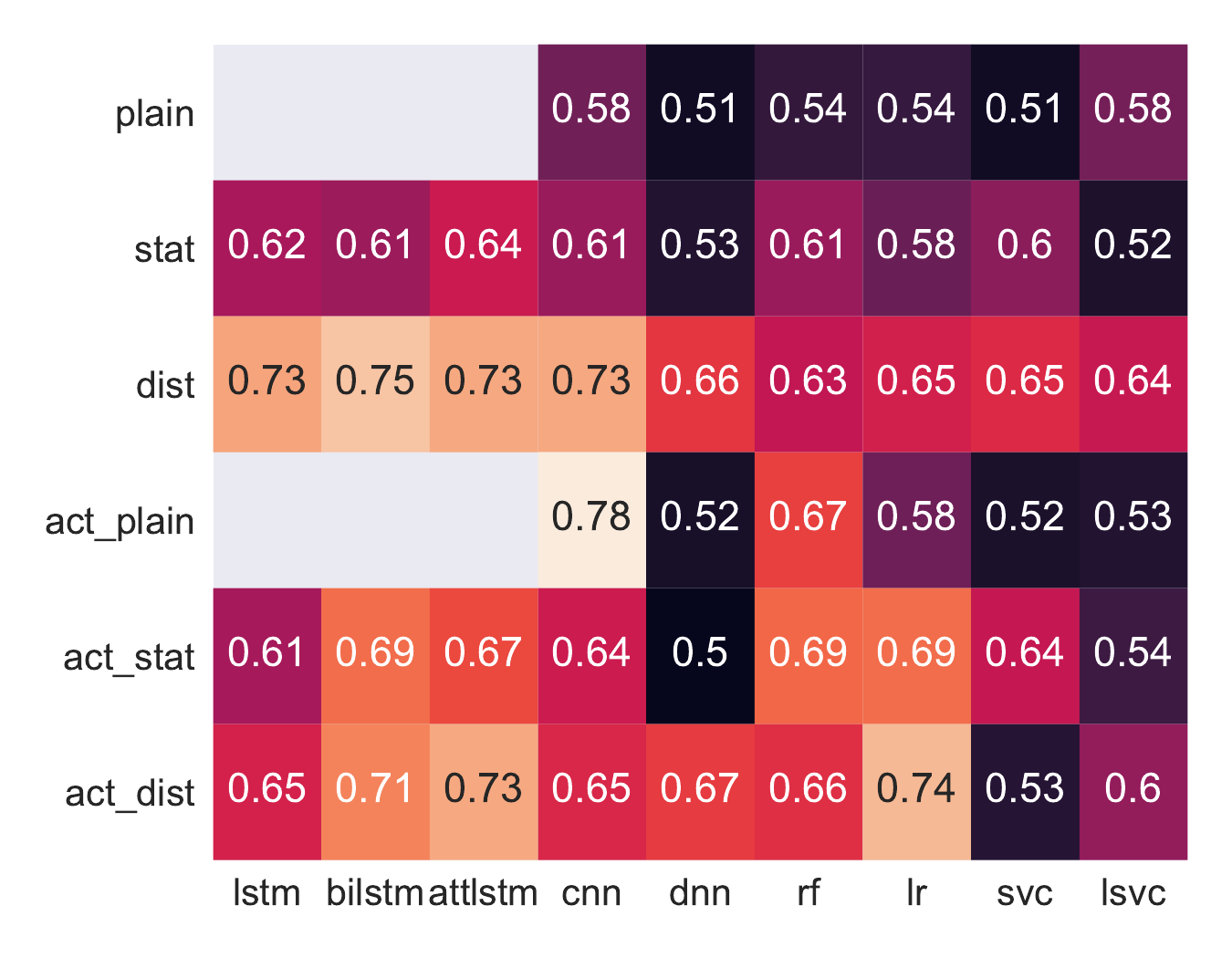}
	\caption{Best AUC scores for gender prediction \label{fig:sex_heat}}
\end{figure}

The results for gender prediction are in Fig.~\ref{fig:sex_heat}. In case of gender prediction, simple statistics are not enough and distributional method seems superior, as it contains more information. Again LSTMs and CNN outperform simpler machine learning methods. On most classifiers we can observe that activities perform better then fixed window size split. This indicates existence of certain walking patterns characteristic for males or females. We do not have data saying what were users doing during those activities, but it would be very interesting to study, which activities expose our gender.
The best result is achieved by a CNN consisting of 2 convolution layers with 6 filters of size 16 each, a 0.5 dropout layer, max pooling layer of pool size 4, dense layer with 100 perceptrons and finally dense layer with 1 perceptron run on plain activities with the prediction for a specific user being an arithmetical mean of predictions of all their activities. The average AUC over 5 fold validated training-test split is 0.780 with standard deviation of 0.024.

\begin{figure}[!b]
	\centering
	\includegraphics[width=\columnwidth]{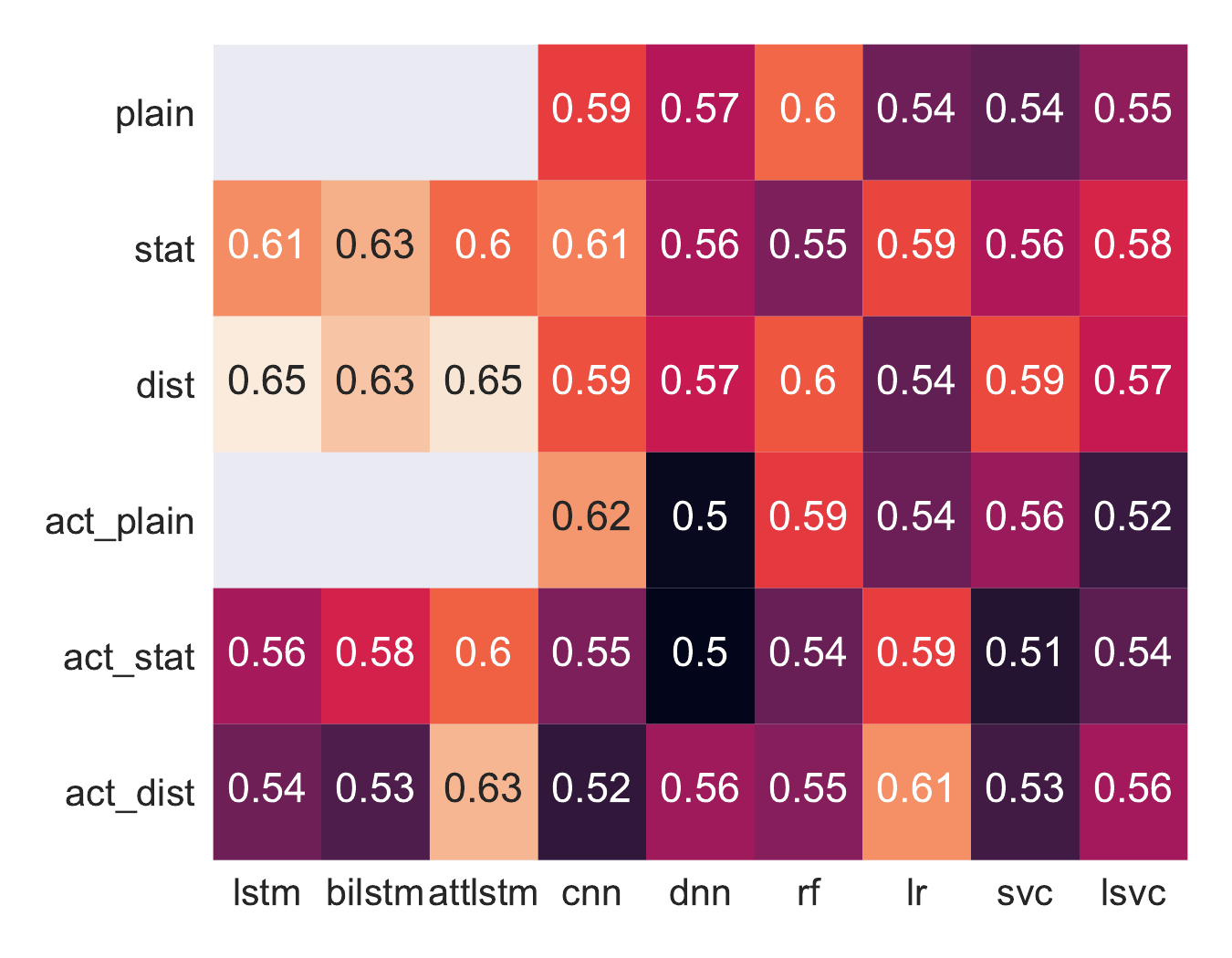}
	\caption{Best AUC scores for education prediction \label{fig:edu_heat}}
\end{figure}

The results for education prediction are in Fig.~\ref{fig:edu_heat}. Education is by far the hardest attribute to predict and we are not able to train a reliable classifier.
LSTMs seem to perform slightly better then other methods, with the best being a standard LSTM with 16 units and a 0.2 dropout layer, trained on distributions with $\bucketsize = 4$ and $\windowsize = 720$ (3h). The average AUC over 5 fold validated training-test split is 0.650 with standard deviation of 0.037.
The reason why we are able to perform better then a random guess is probably because of the correlation of age and education in our data. To verify this claim we use the best performing (for age) CNN classifier, train it on statistical file for the task of predicting age and use it to predict education. In such setting we managed to achieve AUC of 0.637, which confirms our hypothesis.


\section{Linkability}\label{sec:link_attack}

Our linkability attack tries to identify whether two observations of stepcount data belong to the same individual.
In our experiments we focus on daily observations, meaning for each user $\user$ we have 7 daily feature vectors $\vecstepuserday$, where $\dayy \in \week$ and $\week = \{ \mathit{Monday}, \mathit{Tuesday}, \ldots, \mathit{Sunday}\}$.

Formally, let $ \vecstepdailyu^{\,\dayy_1}$ and $\vecstepdailyv^{\, \dayy_2}$, $\dayy_1 \neq \dayy_2$ be the step count feature vectors from two days of users $\user$ and $v$. 
The adversary's objective is to predict the probability that $\user$ and $v$ are the same user, where one or both  are anonymous. 
\[
\Link(\vecstepdailyu^{\,\dayy_1} , \vecstepdailyv^{\, \dayy_2}) = (\user \stackrel{?}{=} v)
\]
$ \dayy_1, \dayy_2 \in \week, \dayy_1 \neq \dayy_2$.
The adversary can use the daily raw step counts per 15s interval $\vecsteprawuserday$, features created by the statistical and distributional methods or their normalized versions as described in sections  \ref{stats_features}, \ref{dist_features} and \ref{data_norm}.

The attack function $\Link$ can be instantiated in three different ways as follows:

\subsection{Similarity based Attack}
This attack is unsupervised and is the simplest one. 
It involves calculating a distance metric between two samples $ \vecstepdailyu^{\,\dayy_1}$ and $\vecstepdailyv^{\,\dayy_2}$ and comparing it to a threshold $t$. If the distance is smaller than $t$, then we predict that the samples came from the same user.
We use the euclidean distance and the cosine distance. 
Other distance metrics have similar or lower performance and are therefore not included. 
For the euclidean distance, the attack function is instantiated as:
\[
\Link_{\eucl}( \vecstepdailyu^{\,\dayy_1}, \vecstepdailyv ^{\,\dayy_1}) = \sqrt{ \sum_{i=1}^{n} (\step_{\user,i}^{\,\dayy_1} - \step_{v,i}^{\,\dayy_2} )^2}  < t_{\eucl}
\]
Similarly, for the cosine distance, the attack function is instantiated as:
\[
\Link_{\cosine}( \vecstepdailyu^{\,\dayy_1}, \vecstepdailyv^{\,\dayy_2} ) = 1 - \frac{\vecstepdailyu^{\,\dayy_1} \vecstepdailyv^{\,\dayy_2 \,T} }{ ||\vecstepdailyu|| ||\vecstepdailyv||} < t_{\cosine}
\]
where $\cdot^{\,T}$ denotes vector transposition and $||\cdot|| $ denotes the $L_2$ norm of the vector. 
$t_{\cosine}$ and $t_{\eucl}$ denotes thresholds for each attack.

Experimenting with different values of thresholds $t_{\cosine}$ and $t_{\eucl}$ gives us a range of false and true positive values. Plotting the false-positive rate on the x-axis and the true-positive rate on the y-axis produces the ROC (Receiver Operating Characteristic) curve. We use the area under this curve, referred to as AUC, to analyze the success of the attack. Unlike other metrics, AUC summarizes the performance in a straightforward manner: 0.5 is as bad as random guessing and 1.0 indicates perfect prediction.
We assume the attacker knows the best threshold $t_{\cos}$ or  $t_{\eucl}$ and therefore get an upper bound for the privacy threat resulting from the unsupervised attack. 
We denote these attacks by $\unsupeucl$ and $\unsupcos$.

\subsection{Random Forest Classifier based Attack}
This attack uses random forest classifier, which is a state-of-the-art supervised machine learning approach. 
We first apply a vector distance metric, namely, L1 distance between the pair of samples.
We then use the resulting distance vectors as features to fit random forest classifiers.
As above, we use the AUC between the positive class probability of the random forest and the true class labels to evaluate the performance of the attack.
L2, element-wise arithmetic mean and Hadamard distance have similar or lower performance and are therefore not included.
We denote this attack by $\RFstandard$.

\subsection{Siamese Neural Network based Attack}

One Shot Learning has become the state of the art solution when large amounts of labelled training data is not available for standard classification with deep neural networks.
This attack uses One Shot Learning with Siamese Neural Networks \cite{bromley1994signature}.  
Siamese networks are a special type of neural network architecture that contain two identical sub-networks that have the same configuration, the same parameters and weights, and mirrored updates during training.
Instead of learning to classify its inputs, this model learns to differentiate between two inputs and finds a similarity or relationship between them.
Specifically, our Siamese Network Model leverages the semantic similarities between pairs of step count vectors of the same user and the differences between pairs of step count vectors of different users.
Just one sample of the true identity of a user in training may be sufficient to predict or recognize this user in the future. 

\begin{figure}[!t]
	\centering
	\includegraphics[width=\columnwidth]{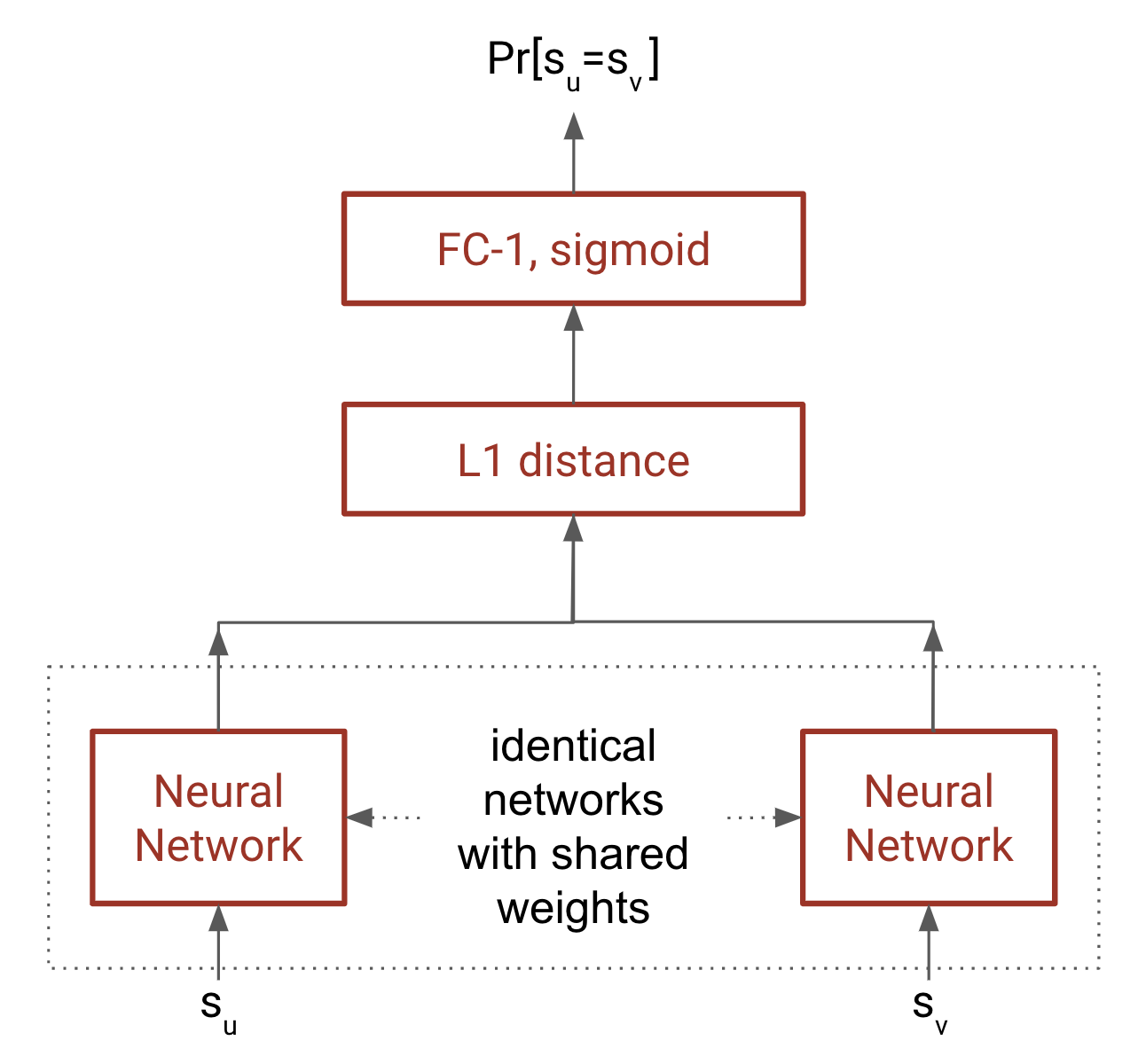}
	\caption{Illustration of a basic Siamese Network\label{fig_siamese}}
\end{figure}

Figure \ref{fig_siamese} shows an illustration of a basic Siamese Neural Network.

We use L1 distance to combine the output of the shared networks, followed by a fully connected layer with sigmoid activation function.
The hypothesis is that two inputs from different users will produce different feature embeddings at the inner layers, and inputs from the same user will result in similar feature embeddings.
Hence the element-wise absolute difference between the two feature embeddings must also be very different between the two cases. 
Therefore the score generated by the output sigmoid layer must also be different in both cases.

This model is similar to having an auto encoder and a distinguishing classifier, but instead we learn encodings useful for this exact purpose of our task. We found that using auto encoders for this task does not produce representations resulting in good predictions and thus we exclude those experiments in the paper. 

We instantiate the two shared subnetworks with different state of the art neural network layers to get five different attack variants as follows:

\subsubsection{Dense Layers}
In the first variant, we use two densely connected layers for the shared subnetworks. The first dense layer contains half the number of neurons as the size of the input.
The second dense layer further contains half the number of neurons as the first one.
We denote this attack by $\densesiamese$.

\subsubsection{LSTM Layers}
In this variant, the shared subnetworks consist of an LSTM layer with 8 units, and dropout of 0.2. We denote this attack by $\lstmsiamese$. 

\subsubsection{Bidirectional LSTM Layers}
In this variant, the shared subnetworks consist of a Bidirectional LSTM layer with 8 units, and dropout of 0.2. We denote this attack by $\bilstmsiamese$. 

\subsubsection{Attention Layers}
In this variant, the shared subnetworks consist of a Luong-style dot-product attention layer, \cite{luong2015effective} applied to the output of the Bidirectional LSTM layer in $\bilstmsiamese$. We denote this attack by $\attntnsiamese$. 

\subsubsection{1D CNN Layers}
In this variant, the shared subnetworks consist of two 1D CNN layers, followed by a max pooling layer of size 8.
Each CNN layer had a filter size of 16 and a kernel size of 6.
The max pooling layer is followed by a flatten layer, which is followed by a densely connected layer of 100 neurons.
The result of using a pooling layer and creating down sampled or pooled feature maps is a summarized version of the features detected in the input. They are useful as small changes in the location of the feature in the input detected by the convolutional layer will result in a pooled feature map with the feature in the same location.
We denote this attack by $\cnnsiamese$.

As before, we calculate the AUC between the true class labels and the result of the sigmoid function to evaluate the success of the attack.

\section{Linkability - Evaluation} \label{sec:link_eva}
We now present the experimental evaluation of our linkability attacks, namely our  unsupervised attacks $\unsupeucl$ and $\unsupcos$, our random forest attack $\RFstandard$ and our three Siamese network attacks $\cnnsiamese$,  $\lstmsiamese$ and $\densesiamese$ on our dataset.

\subsection{Experimental Setup}
We evaluate our linkability attacks on various stepcount features $\vecstepuserday$ which include raw step counts $\vecsteprawuserday$ as well as $\vecstepstatuserday, \mathit{stat} \in \{ \mathit{sum},   \mathit{max},  \mathit{median},  \mathit{mean}\}$ and distributions $\vecstepdistuser^{\,\dayy}$, over a period of 24 hours on a day $\dayy \in \week$.
We also evaluate the attacks on the normalized (feature-wise normalization) versions of each feature as described in \ref{norm} and denote the results by the suffix $\mathtt{\_norm}$. 
For example we denote the results of the attack $\unsupeucl$ on the normalized features by $\unsupeucl\_\mathtt{norm}$. We remove features with variance less than $0.001$ before fitting into each model.

We create all possible pairs of daily step counts $\vecstepuserday$ out of 7 days of data for each user to obtain positive class samples (vectors representing two different days of the same user). Thus for each user $\user$ we have $\binom{7}{2}$, i.e. 21 positive samples in the form $\{ \vecstepdailyu^{\,\dayy_1} , \vecstepdailyu^{\, \dayy_2}\},\:  \dayy_1, \dayy_2 \in  \week,\:  \dayy_1 \neq \dayy_2$.
For negative samples, we randomly pick an equal number of pairs of daily step counts from different users in the form $\{ \vecstepdailyu^{\,\dayy_1} , \vecstepdailyv^{\, \dayy_2}\},\:  \dayy_1, \dayy_2 \in  \week,\:  \user \neq v$.

We perform a 5-fold cross validation to evaluate our attacks.
The positive and negative samples are equally divided into both sets.
We do not optimize any parameters for the supervised attacks.
The results of our supervised attacks are therefore a lower bound on the privacy risk arising out of such an adversary model.

Following the above strategies and removing symmetric user pairs, we get 20937 samples in each class and 41874 samples in total. We train on 33500 samples, test on 8374 samples in each iteration. 

\subsection{Results}

\begin{figure}[!ht]
	\centering
	\includegraphics[width=\columnwidth]{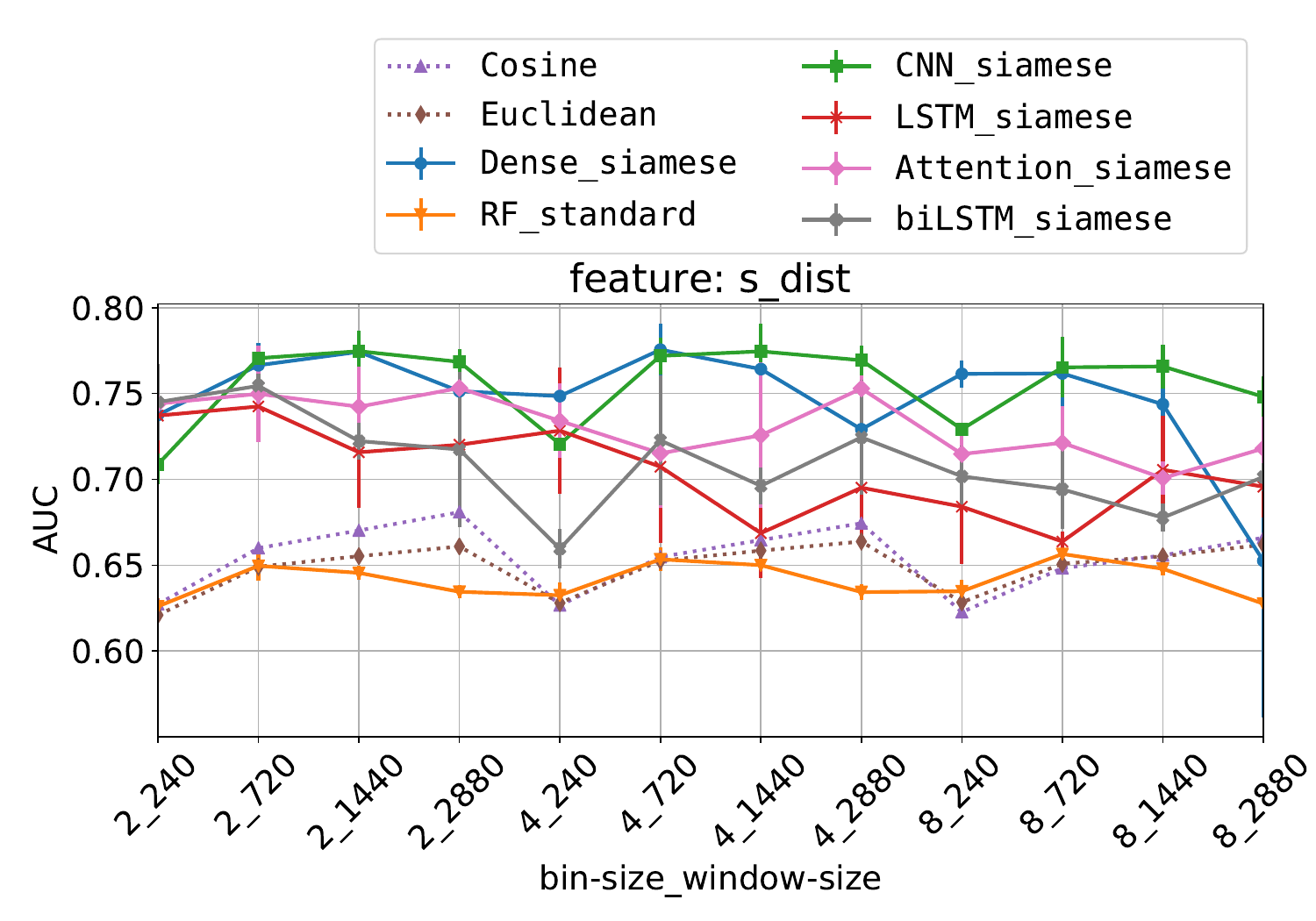}
	\caption{Performance of all attacks on distribution feature $\vecstep_{\mathit{dist}}$. The x-axis indicates the bucket size \bucketsize followed by the window size \windowsize. \label{dist_plot}}
\end{figure}

Figure \ref{dist_plot} shows the results of our attacks on the features $\vecstep_{\mathit{dist}}$. The y-axis indicates the AUC (mean and the standard deviation via an error bar over all cross validation folds). The x-axis indicates different bucket size $\bucketsize$ and window size $\windowsize$ combinations for the distributions in the input file. 
We notice that the unsupervised attacks $\unsupeucl$ and $\unsupcos$ and the standard random forest classifier $\RFstandard$ have very low performance. 
However our $\densesiamese$ and $\cnnsiamese$ classifiers hugely outperform them and achieves AUC higher than 0.75 for most of the inputs. The RNN based classifiers have only a slightly lower performance; the $\lstmsiamese$ is outperformed by the $\bilstmsiamese$, which is further outperformed by the $\attntnsiamese$ as expected.
For big bucket and window sizes, we have lower number of features. Therefore the $\densesiamese$ attack does not have a significant advantage over the simpler attacks. We notice that its performance is almost the same as the unsupervised attacks for $\bucketsize\_\windowsize = 8\_2880$.

\begin{figure}[!ht]
	\centering
	\includegraphics[width=\columnwidth]{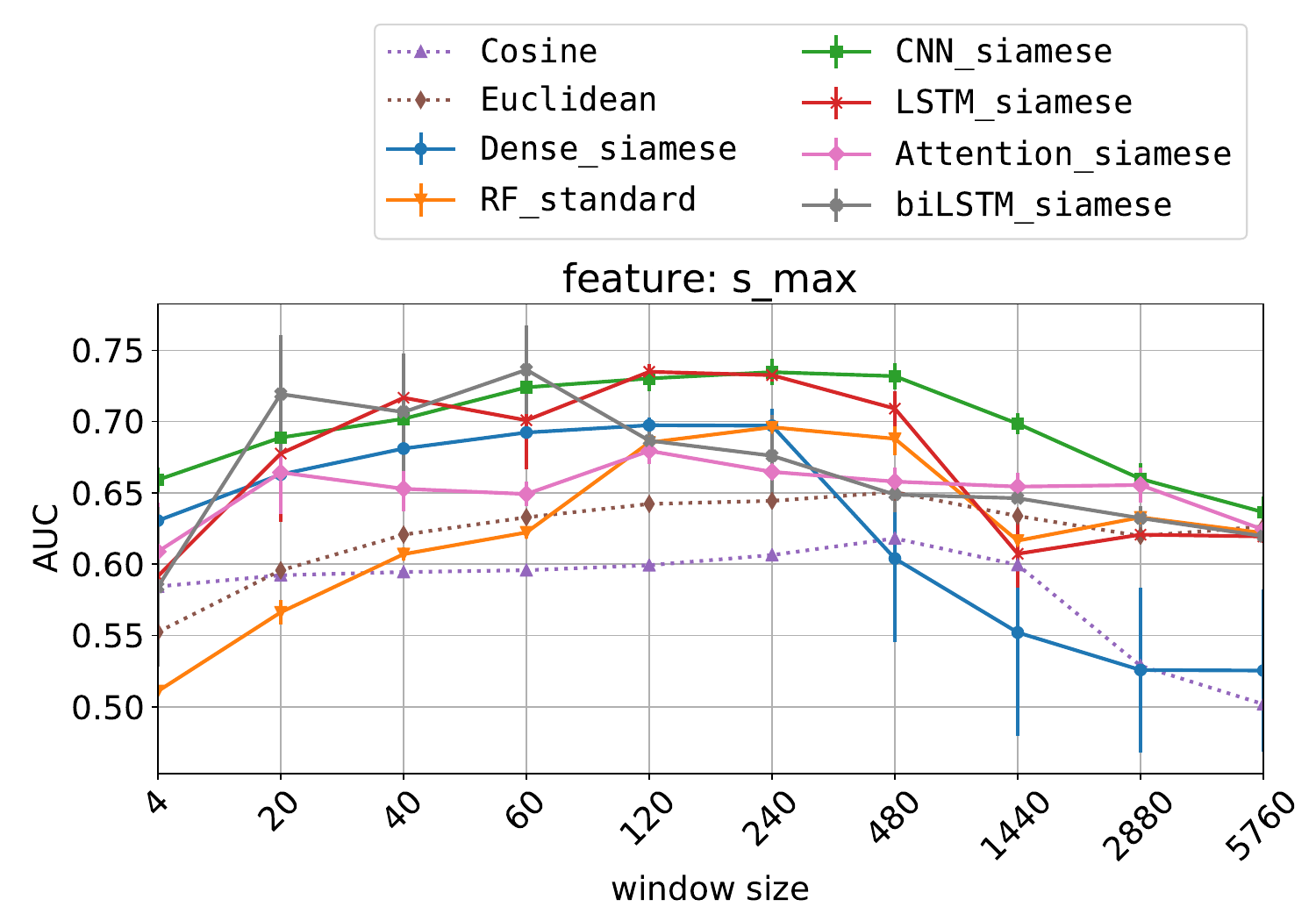}
	\caption{Performance of all attacks on the statistical feature $\vecstep_{\mathit{max}}$ \label{max_plot}}
\end{figure}

\begin{figure}[!ht]
	\centering
	\includegraphics[width=\columnwidth]{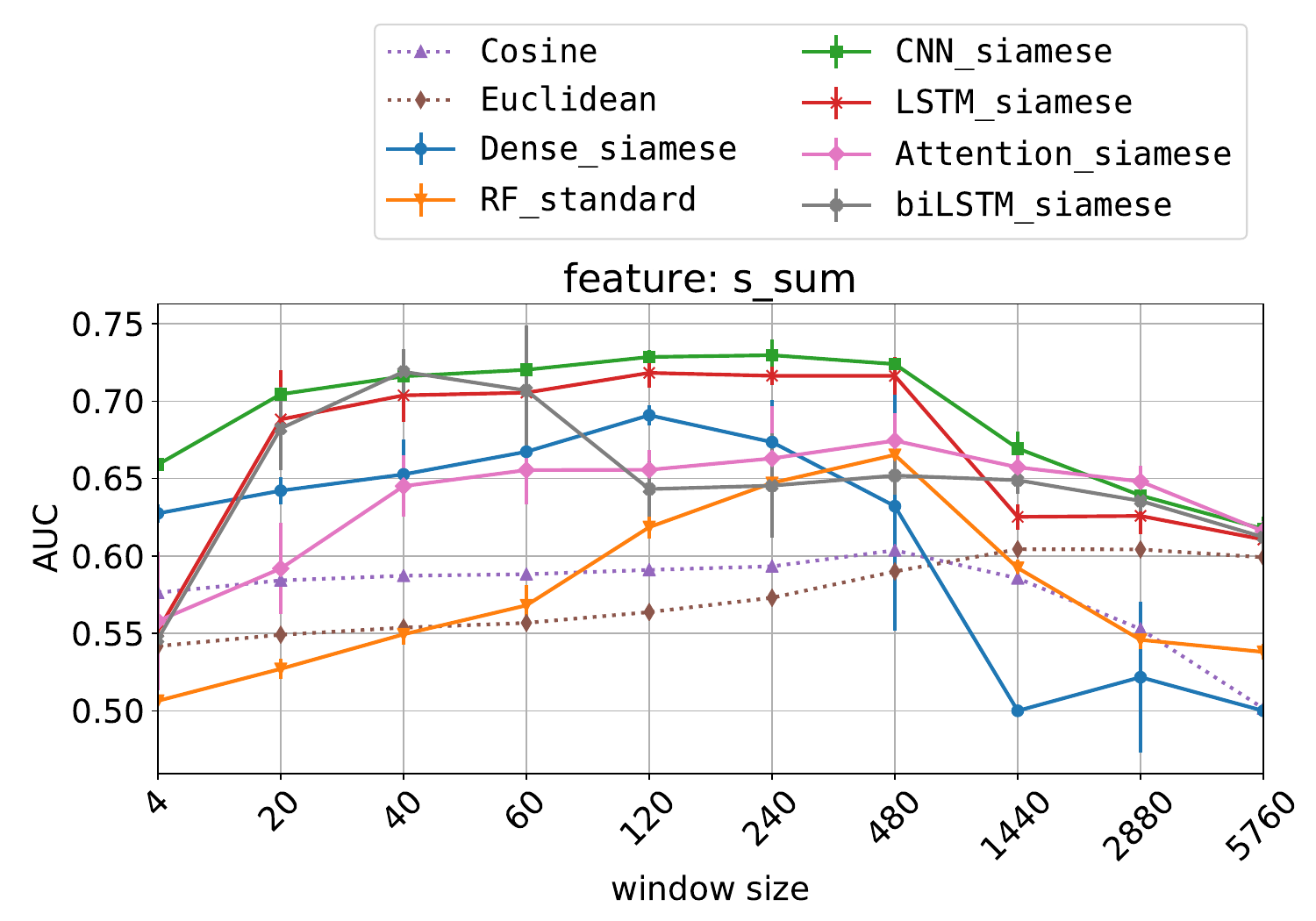}
	\caption{Performance of all attacks on statistical feature  $\vecstep_{\mathit{sum}}$ \label{sum_plot}}
\end{figure}

\begin{figure}[!ht]
	\centering
	\includegraphics[width=\columnwidth]{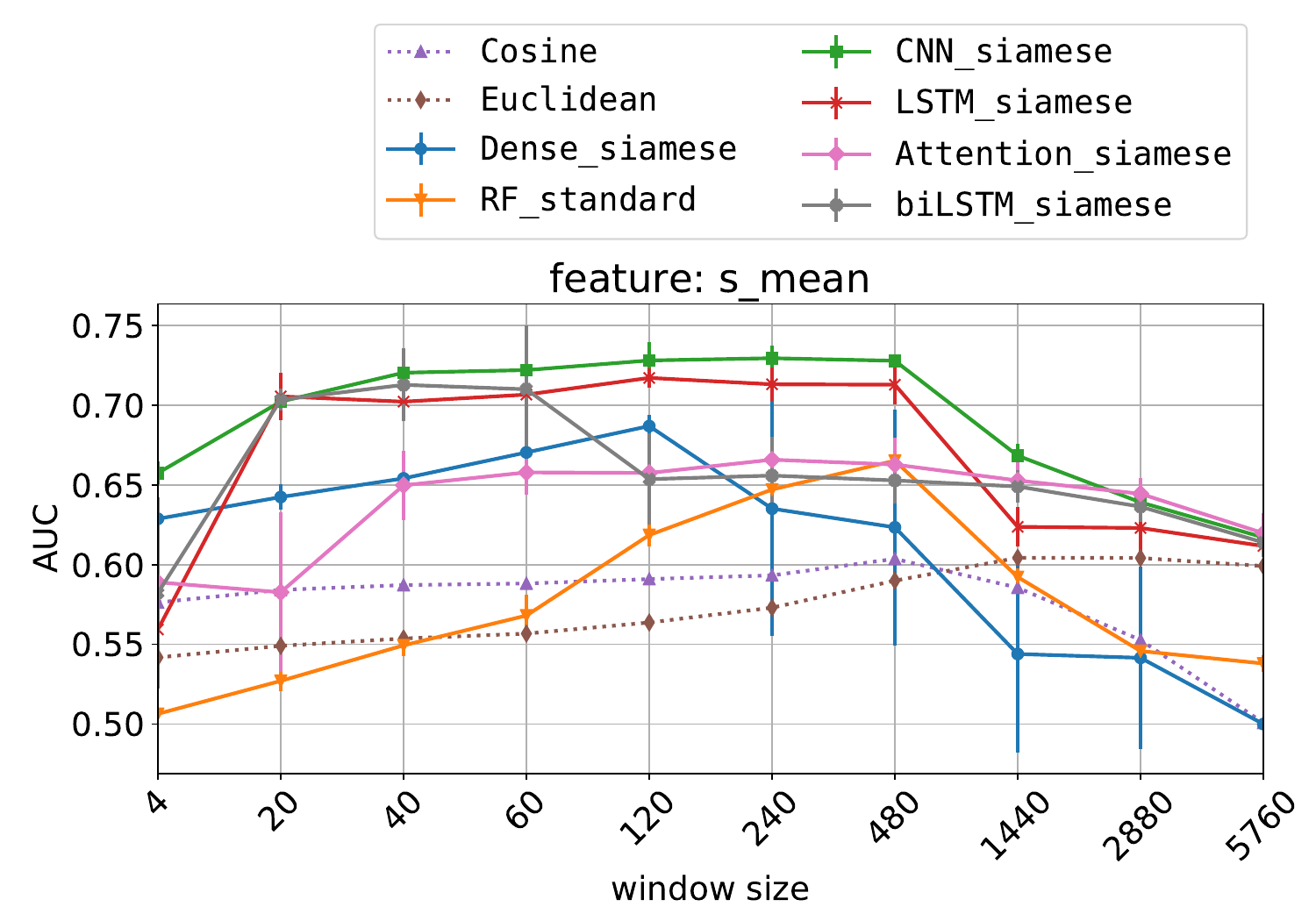}
	\caption{Performance of all attacks on statistical feature $\vecstep_{\mathit{mean}}$ \label{mean_plot}}
\end{figure}

\begin{figure}[!ht]
	\centering
	\includegraphics[width=\columnwidth]{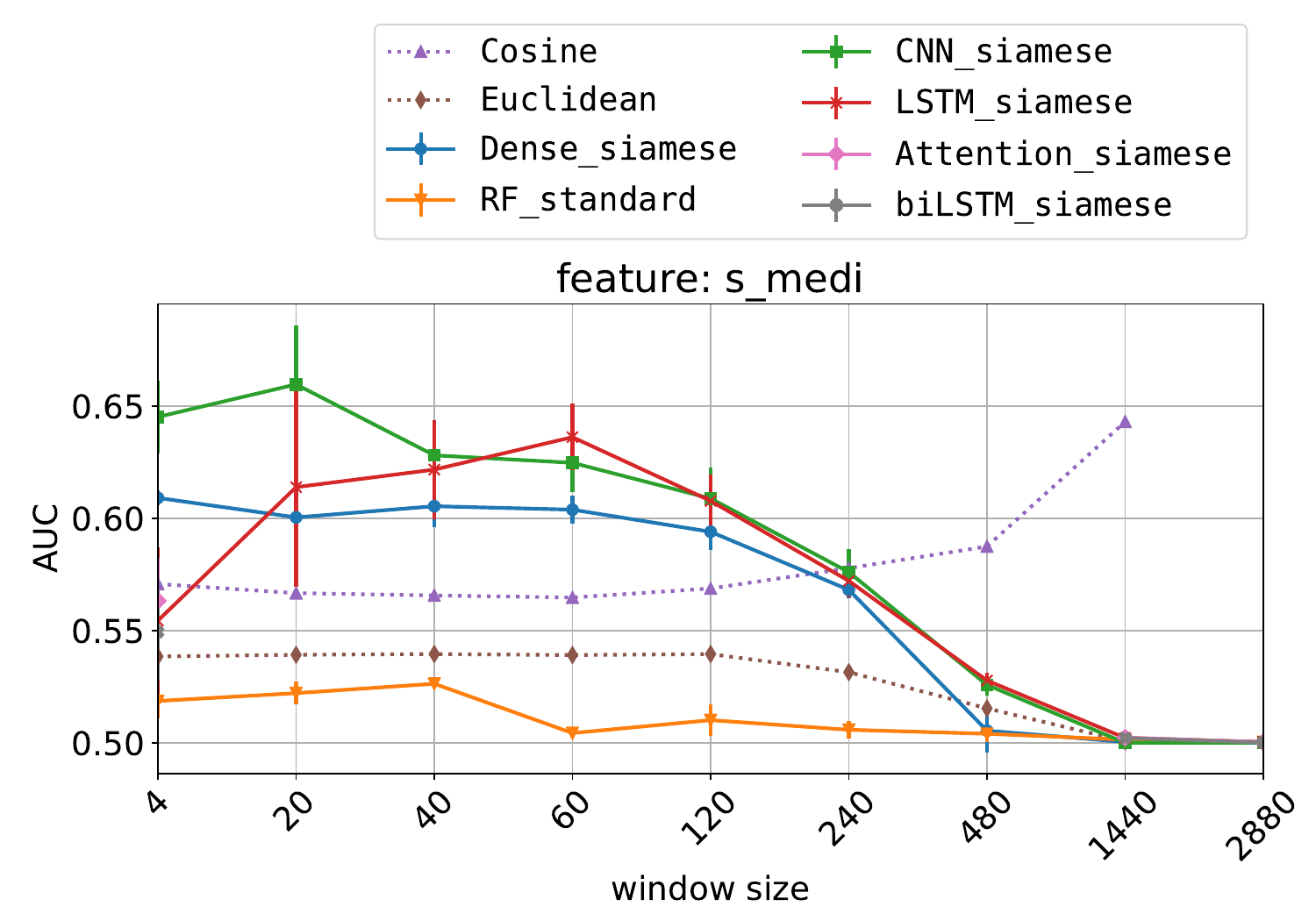}
	\caption{Performance of all attacks on statistical feature  $\vecstep_{\mathit{medi}}$ \label{medi_plot}}
\end{figure}

Figures \ref{max_plot}-\ref{medi_plot} shows the results of our attacks on the features $\vecstep_{\mathit{max}}$, $\vecstep_{\mathit{sum}}$, $\vecstep_{\mathit{mean}}$ and $\vecstep_{\mathit{medi}}$.
The x-axis shows increasing window size $\windowsize$ over which the statistic is calculated.
We observe that our neural network based siamese attacks outperform the  $\RFstandard$, $\unsupeucl$ and $\unsupcos$ for smaller window sizes. However as window size increases, the performance of the siamese attacks drops quite fast.
We also observe that $\cnnsiamese$ and $\bilstmsiamese$  always achieved the top AUCs for each of the 5 feature extraction methods, closely followed by $\lstmsiamese$.
The $\attntnsiamese$ struggles for small window sizes (larger number of timesteps), however as the number of timesteps increase, it outperforms $\bilstmsiamese$ and $\lstmsiamese$.
This shows that the success of attention mechanism is limited to sequences of smaller length.

\begin{figure}[!ht]
	\centering
	\includegraphics[width=\columnwidth]{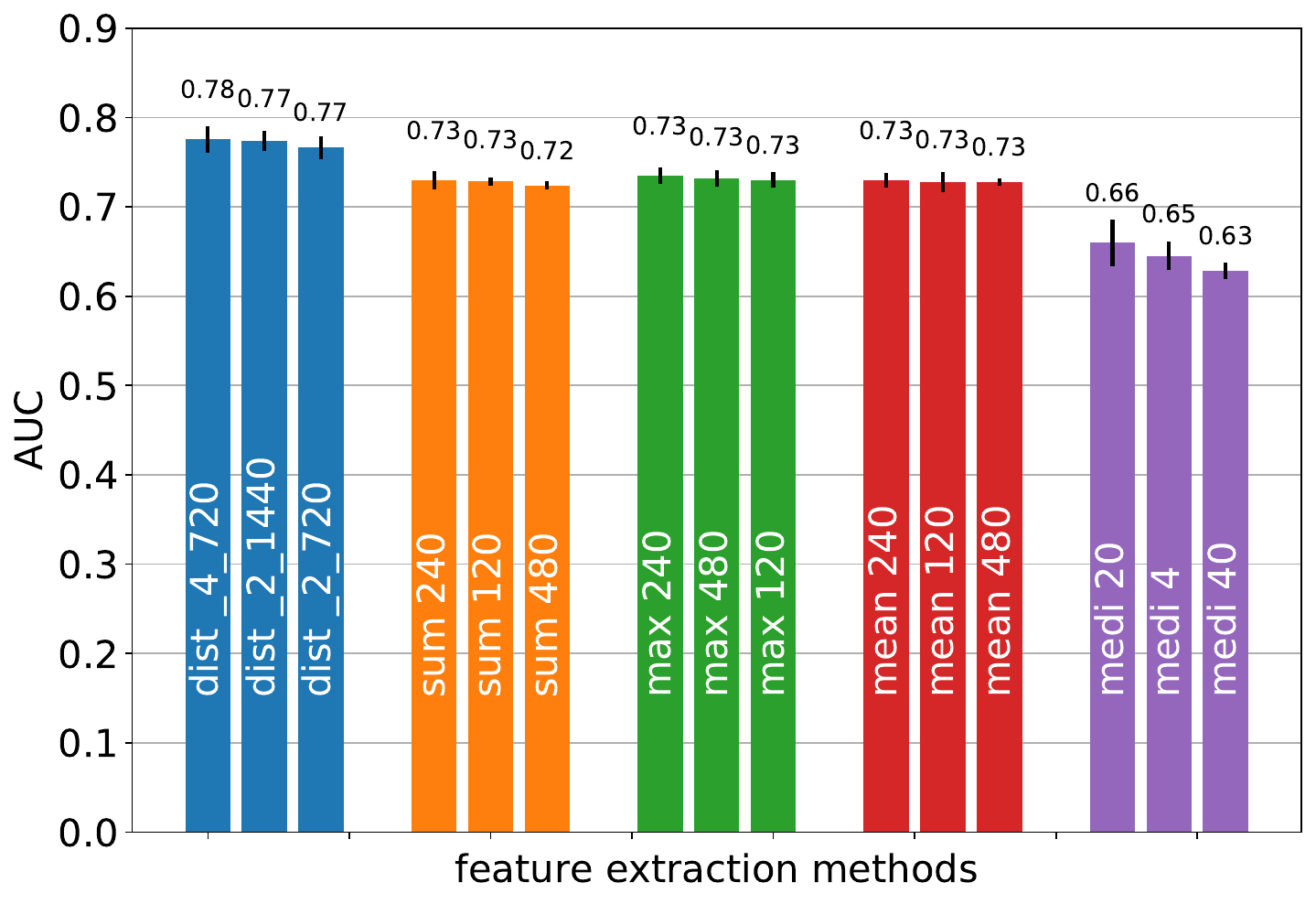}
	\caption{Top 3 AUCs achieved by input features from each feature extraction method (with the $\densesiamese$ classifier).  \label{top3}}
\end{figure}

We further compare the five different feature extraction methods with each other to see which captured the most information for user linkability.
Figure \ref{top3} shows the AUCs (mean and standard deviation over all cross validation folds) achieved by the three best performing inputs of each feature extraction method. We see that features extracted by the distribution method $\vecstep_{\mathit{dist}}$ outperform the rest and as expected $\vecstep_{\mathit{medi}}$, i.e. the median statistic shows the poorest performances. The highest average AUC, achieved by $\vecstep_{\mathit{dist}}$ with $\bucketsize\_\windowsize = 4\_720$, is 0.78 using $\densesiamese$. 

\clearpage
\section{Conclusion}\label{sec:conclusion}
We perform the first systematic analysis of privacy risks arising from physical activity data. 
Via an extensive evaluation on a real life dataset, we demonstrate that indeed step count data poses significant threat to various aspects of users' privacy.

With a relatively small dataset of 1000 users only and using simple machine learning classifiers, we find that an adversary having access to actimetry (step count) data of different users can perform a linkability attack with a high confidence.
Attribute inference is harder, but still possible. 
Among all three attributes, prediction of age is the easiest, while education cannot be predicted reliably.
For gender inference specifically, we obtain promising results by averaging predictions from classifying shorter walking periods (activities). Some of these short patterns strongly indicate users' gender but not education or age. Exploring such smaller patterns for different attributes would be an interesting direction for future work. 
Note that our goal is not to quantify the worst case privacy risk of a specific attack but to explore a wide range of different feature extraction methods and classifiers in a spectrum of privacy attacks.

Actimetry data is not currently regarded as privacy sensitive, but big players, having access to an even bigger dataset could deanonymize users and infer their hidden attributes much easier.
We would like to bring the attention of policymakers to the sensitivity of pedometer data and propose that users should be made aware of the risks of sharing their step count data.
This highlights the need for further research on privacy preserving ways of collecting such data.


\bibliographystyle{IEEEtranS}
\bibliography{biblio}

\begin{thebibliography}{10}
\providecommand{\url}[1]{#1}
\csname url@samestyle\endcsname
\providecommand{\newblock}{\relax}
\providecommand{\bibinfo}[2]{#2}
\providecommand{\BIBentrySTDinterwordspacing}{\spaceskip=0pt\relax}
\providecommand{\BIBentryALTinterwordstretchfactor}{4}
\providecommand{\BIBentryALTinterwordspacing}{\spaceskip=\fontdimen2\font plus
\BIBentryALTinterwordstretchfactor\fontdimen3\font minus
  \fontdimen4\font\relax}
\providecommand{\BIBforeignlanguage}[2]{{%
\expandafter\ifx\csname l@#1\endcsname\relax
\typeout{** WARNING: IEEEtranS.bst: No hyphenation pattern has been}%
\typeout{** loaded for the language `#1'. Using the pattern for}%
\typeout{** the default language instead.}%
\else
\language=\csname l@#1\endcsname
\fi
#2}}
\providecommand{\BIBdecl}{\relax}
\BIBdecl

\bibitem{aguiar2014monitoring}
B.~Aguiar, J.~Silva, T.~Rocha, S.~Carneiro, and I.~Sousa, ``Monitoring physical
  activity and energy expenditure with smartphones,'' in \emph{2014 IEEE-EMBS
  International Conference on Biomedical and Health Informatics, BHI 2014}, 06
  2014, pp. 664--667.

\bibitem{al2012homophily}
F.~Al~Zamal, W.~Liu, and D.~Ruths, ``Homophily and latent attribute inference:
  Inferring latent attributes of twitter users from neighbors,'' in \emph{Sixth
  International AAAI Conference on Weblogs and Social Media}, 2012.

\bibitem{althoff2017large}
T.~Althoff, J.~L. Hicks, A.~C. King, S.~L. Delp, J.~Leskovec \emph{et~al.},
  ``Large-scale physical activity data reveal worldwide activity inequality,''
  \emph{Nature}, vol. 547, no. 7663, pp. 336--339, 2017.

\bibitem{bittel2015accuracy}
A.~Bittel, A.~Elazzazi, and D.~Bittel, ``Accuracy and precision of an
  accelerometer-based smartphone app designed to monitor and record angular
  movement over time,'' \emph{Telemedicine journal and e-health : the official
  journal of the American Telemedicine Association}, vol.~22, 10 2015.

\bibitem{BojinovMNB14}
\BIBentryALTinterwordspacing
H.~Bojinov, Y.~Michalevsky, G.~Nakibly, and D.~Boneh, ``Mobile device
  identification via sensor fingerprinting,'' \emph{CoRR}, vol. abs/1408.1416,
  2014. [Online]. Available: \url{http://arxiv.org/abs/1408.1416}
\BIBentrySTDinterwordspacing

\bibitem{bromley1994signature}
J.~Bromley, I.~Guyon, Y.~LeCun, E.~S{\"a}ckinger, and R.~Shah, ``Signature
  verification using a" siamese" time delay neural network,'' in \emph{Advances
  in neural information processing systems}, 1994, pp. 737--744.

\bibitem{ciot2013gender}
M.~Ciot, M.~Sonderegger, and D.~Ruths, ``Gender inference of twitter users in
  non-english contexts,'' in \emph{Proceedings of the 2013 Conference on
  Empirical Methods in Natural Language Processing}, 2013, pp. 1136--1145.

\bibitem{davarci2017age}
E.~{Davarci}, B.~{Soysal}, I.~{Erguler}, S.~O. {Aydin}, O.~{Dincer}, and
  E.~{Anarim}, ``Age group detection using smartphone motion sensors,'' in
  \emph{2017 25th European Signal Processing Conference (EUSIPCO)}, 2017, pp.
  2201--2205.

\bibitem{Sanorita2014accelprint}
\BIBentryALTinterwordspacing
S.~Dey, N.~Roy, W.~Xu, R.~R. Choudhury, and S.~Nelakuditi, ``Accelprint:
  Imperfections of accelerometers make smartphones trackable,'' in \emph{21st
  Annual Network and Distributed System Security Symposium, {NDSS} 2014, San
  Diego, California, USA, February 23-26, 2014}.\hskip 1em plus 0.5em minus
  0.4em\relax The Internet Society, 2014. [Online]. Available:
  \url{https://www.ndss-symposium.org/ndss2014/accelprint-imperfections-accelerometers-make-smartphones-trackable}
\BIBentrySTDinterwordspacing

\bibitem{dong2014demographics}
\BIBentryALTinterwordspacing
Y.~Dong, Y.~Yang, J.~Tang, Y.~Yang, and N.~V. Chawla, ``Inferring user
  demographics and social strategies in mobile social networks,'' in
  \emph{Proceedings of the 20th ACM SIGKDD International Conference on
  Knowledge Discovery and Data Mining}, ser. KDD '14.\hskip 1em plus 0.5em
  minus 0.4em\relax New York, NY, USA: Association for Computing Machinery,
  2014, p. 15–24. [Online]. Available:
  \url{https://doi.org/10.1145/2623330.2623703}
\BIBentrySTDinterwordspacing

\bibitem{gl2018attribute}
\BIBentryALTinterwordspacing
N.~Z. Gong and B.~Liu, ``Attribute inference attacks in online social
  networks,'' \emph{ACM Trans. Priv. Secur.}, vol.~21, no.~1, Jan. 2018.
  [Online]. Available: \url{https://doi.org/10.1145/3154793}
\BIBentrySTDinterwordspacing

\bibitem{guidoux2014activity}
R.~Guidoux, M.~Duclos, G.~Fleury, P.~Lacomme, N.~Lamaudière, P.-H. Manenq,
  L.~Paris, L.~Ren, and S.~Rousset, ``A smartphone-driven methodology for
  estimating physical activities and energy expenditure in free living
  conditions,'' \emph{Journal of Biomedical Informatics}, vol.~52, 07 2014.

\bibitem{hassan2018analysis}
W.~U. Hassan, S.~Hussain, and A.~Bates, ``Analysis of privacy protections in
  fitness tracking social networks-or-you can run, but can you hide?'' in
  \emph{27th $\{$USENIX$\}$ Security Symposium ($\{$USENIX$\}$ Security 18)},
  2018, pp. 497--512.

\bibitem{hu2007demographics}
\BIBentryALTinterwordspacing
J.~Hu, H.-J. Zeng, H.~Li, C.~Niu, and Z.~Chen, ``Demographic prediction based
  on user's browsing behavior,'' in \emph{Proceedings of the 16th International
  Conference on World Wide Web}, ser. WWW '07.\hskip 1em plus 0.5em minus
  0.4em\relax New York, NY, USA: Association for Computing Machinery, 2007, p.
  151–160. [Online]. Available: \url{https://doi.org/10.1145/1242572.1242594}
\BIBentrySTDinterwordspacing

\bibitem{jiang2019pedometer}
Y.~{Jiang}, W.~{Tang}, N.~{Gao}, J.~{Xiang}, D.~{Zha}, and X.~{Li}, ``Your
  pedometer tells you: Attribute inference via daily walking step count,'' in
  \emph{2019 IEEE SmartWorld, Ubiquitous Intelligence Computing, Advanced
  Trusted Computing, Scalable Computing Communications, Cloud Big Data
  Computing, Internet of People and Smart City Innovation
  (SmartWorld/SCALCOM/UIC/ATC/CBDCom/IOP/SCI)}, 2019, pp. 834--842.

\bibitem{jiang2020attr}
\BIBentryALTinterwordspacing
Y.~Jiang, W.~Tang, N.~Gao, J.~Xiang, C.~Tu, and M.~Li, ``Multiple demographic
  attributes prediction in mobile and sensor devices,'' in \emph{Advances in
  Knowledge Discovery and Data Mining - 24th Pacific-Asia Conference, {PAKDD}
  2020, Singapore, May 11-14, 2020, Proceedings, Part {I}}, ser. Lecture Notes
  in Computer Science, H.~W. Lauw, R.~C. Wong, A.~Ntoulas, E.~Lim, S.~Ng, and
  S.~J. Pan, Eds., vol. 12084.\hskip 1em plus 0.5em minus 0.4em\relax Springer,
  2020, pp. 857--868. [Online]. Available:
  \url{https://doi.org/10.1007/978-3-030-47426-3\_66}
\BIBentrySTDinterwordspacing

\bibitem{kalimeri2019demographics}
\BIBentryALTinterwordspacing
K.~Kalimeri, M.~G. Beiró, M.~Delfino, R.~Raleigh, and C.~Cattuto, ``Predicting
  demographics, moral foundations, and human values from digital behaviours,''
  \emph{Computers in Human Behavior}, vol.~92, p. 428–445, Mar 2019.
  [Online]. Available: \url{http://dx.doi.org/10.1016/j.chb.2018.11.024}
\BIBentrySTDinterwordspacing

\bibitem{luong2015effective}
M.-T. Luong, H.~Pham, and C.~D. Manning, ``Effective approaches to
  attention-based neural machine translation,'' \emph{arXiv preprint
  arXiv:1508.04025}, 2015.

\bibitem{meteriz2019you}
{\"U}.~Meteriz, N.~F. Y{\i}ld{\i}ran, and A.~Mohaisen, ``You can run, but you
  cannot hide: Using elevation profiles to breach location privacy through
  trajectory prediction,'' \emph{arXiv preprint arXiv:1910.09041}, 2019.

\bibitem{nguyen2019location}
K.~A. Nguyen, R.~N. Akram, K.~Markantonakis, Z.~Luo, and C.~Watkins, ``Location
  tracking using smartphone accelerometer and magnetometer traces,'' in
  \emph{Proceedings of the 14th International Conference on Availability,
  Reliability and Security}, 2019, pp. 1--9.

\bibitem{pei2013behavior}
L.~Pei, R.~Guinness, R.~Chen, J.~Liu, H.~Kuusniemi, Y.~Chen, L.~Chen, and
  J.~Kaistinen, ``Human behavior cognition using smartphone sensors,''
  \emph{Sensors}, vol.~13, pp. 1402--1424, 02 2013.

\bibitem{pierson2018modeling}
E.~Pierson, T.~Althoff, and J.~Leskovec, ``Modeling individual cyclic variation
  in human behavior,'' in \emph{Proceedings of the 2018 World Wide Web
  Conference}, 2018, pp. 107--116.

\bibitem{shameli2017gamification}
A.~Shameli, T.~Althoff, A.~Saberi, and J.~Leskovec, ``How gamification affects
  physical activity: Large-scale analysis of walking challenges in a mobile
  application,'' in \emph{Proceedings of the 26th International Conference on
  World Wide Web Companion}, 2017, pp. 455--463.

\bibitem{vitak2018privacy}
J.~Vitak, Y.~Liao, P.~Kumar, M.~Zimmer, and K.~Kritikos, ``Privacy attitudes
  and data valuation among fitness tracker users,'' in \emph{International
  Conference on Information}.\hskip 1em plus 0.5em minus 0.4em\relax Springer,
  2018, pp. 229--239.

\end{thebibliography}

\end{document}